\documentclass[12pt,english]{article}
\usepackage[T1]{fontenc}
\usepackage[latin9]{inputenc}
\usepackage{geometry}
\usepackage{xcolor}
\usepackage{array}
\usepackage{pifont}
\usepackage{float}
\usepackage{wrapfig}
\usepackage{textcomp}
\usepackage{mathrsfs}
\usepackage{amsbsy}
\usepackage{graphicx}

\makeatletter

\providecommand{\tabularnewline}{\\}

\newcommand{\lyxaddress}[1]{
	\par {\raggedright #1
	\vspace{1.4em}
	\noindent\par}
}

\date{}

\AtBeginDocument{
  
}

\makeatother

\usepackage{babel}
\begin{document}
\title{Stochastic gene transcription with non-competitive transcription regulatory
architecture}
\author{Amit Kumar Das$^{\#}$\thanks{mr.das201718@yahoo.com} }
\maketitle

\lyxaddress{$^{\#}$Kharial High School, Kanaipur, Hooghly-712234, India.}
\begin{abstract}
The transcription factors, such as activators and repressors, can
interact with the promoter of gene either in a competitive or non-competitive
way. In this paper, we construct a stochastic model with non-competitive
transcriptional regulatory architecture and develop an analytical
theory that re-establishes the experimental results with an improved
data fitting. The analytical expressions in the theory allow us to
study the nature of the system corresponding to any of its parameters,
and hence enable us to find out the factors that govern the regulation
of gene expression for that architecture. We notice that, along with
transcriptional reinitiation and repressors, there are other parameters
that can control the noisiness of this network. We also observe that,
the Fano factor (at mRNA level) varies from sub-Poissonian regime
to super-Poissonian regime. In addition to the aforementioned properties,
we observe some anomalous characteristics of the Fano factor (at mRNA
level) and that of the variance of protein at lower activator concentrations
in presence of repressor molecules. This model is useful to understand
the architecture of interactions which may buffer the stochasticity
inherent to gene transcription.
\end{abstract}

\section{Introduction}

In the last two decades, it has been established experimentally that
gene expression and its regulation, a fundamental cellular process
whereby the functional protein molecules are produced in cells, are
inherently stochastic processes \cite{key-01,key-02,key-03,key-04,key-05,key-06,key-07,key-08,key-09,key-10,key-11}.
Along with the experimental works, many theoretical analyses, especially
with exact analytical results, have uplifted the field to a new height
and made the field more fascinating and challenging \cite{key-12,key-13,key-14,key-15,key-16,key-17,key-42,key-43,key-44,key-45,key-46}. 

Gene expression and its regulation are of fundamental importance in
living organisms. They consist of several complex stochastic events
such as, transcription, translation, degradation, etc. \cite{key-39,key-40}.
Transcriptional regulation \cite{key-41,key-42,key-43} plays an essential
role in the development, complexity, and homeostasis of all organisms,
as transcription is the first step of biological information transformation
from genome to proteome. Regulation of transcription is a result of
the interactions between the promoter of gene and regulatory proteins
called the transcription factors (TFs). TFs are classified, according
to their function, as activators and repressors. The activator and
repressor molecules are actively involved in the regulation of gene
transcription, both in prokaryotes and eukaryotes \cite{key-18,key-29,key-31}.
Transcriptional repressors, such as lac and tryptophan repressors,
are well known for prokaryotic systems. Repressor molecules inhibit
the gene transcription by binding to the appropriate region of the
promoter. In comparison to the prokaryotic systems, eukaryotic systems
are much more complex and have compact chromatin structures. For the
initiation of transcription in eukaryotes, remodeling of the chromatin
structure is essential so that the transcription factors and the RNA
polymerase (RNAP) have access to the appropriate binding regions of
the promoter. Thus, gene activation in the eukaryotic system means
the relief of repression by the nucleosomal structure of the chromatin
before the binding of activators \cite{key-51}. Activator and repressor
protein concentrations can be varied by varying the inducer molecules
such as galactose (GAL), aTc (anhydrotetracycline), doxycycline (dox),
etc. \cite{key-03,key-10,key-11}. 

In eukaryotes, regulation of transcription by any of the TFs is modeled
by a two-state telegraphic process \cite{key-14,key-15,key-17}. In
that model, the gene can be either in the ON/active or OFF/inactive
state depending on whether the TFs are bound to the gene or not \cite{key-05,key-36}.
From the active state of the gene, a burst of mRNAs is produced randomly.
The random burst of mRNA synthesis interspersed with a long period
of inactivity is the most important source of cellular heterogeneity
\cite{key-07,key-15,key-16}. However, the causes and consequences
of transcriptional bursts are still very little known. It has not
been possible to view the transcriptional activity of a single gene
in a living eukaryotic cell. It is therefore unclear how long and
how frequently a gene is actively transcribed.

In the burst model or two-state telegraphic model of gene expression,
the initiation of transcription by the recruitment of RNAP II at the
activated state of the promoter is ignored. The first step in transcription
initiation is the recruitment of RNAP II and other transcription machinery
to the promoter to form a pre-initiation complex. After initiation,
a subset of the transcription machinery in the pre-initiation complex
dissociates from the promoter and RNAP II moves forward to transcribe
the gene (polymerase pause release \cite{key-27,key-50}). To begin
the second round of transcription, also called the reinitiation \cite{key-34},
this subset of the transcription machinery along with the RNAP II
must again be recruited to the promoter. It has been shown both experimentally
\cite{key-03,key-06,key-09,key-23,key-24,key-25} and theoretically
\cite{key-26,key-27,key-28}, that reinitiation of transcription by
RNAP II can be crucial for cellular heterogeneity.

The origin and consequences of cellular heterogeneity due to transcriptional
regulation by activators and/or repressors, along with the reinitiation
of transcription by RNAP II, becomes increasingly important. Blake
\textit{et al.} have studied a synthetic GAL1$^{*}$ promoter in yeast
\cite{key-03,key-49}. The transcriptional regulation of the yeast
GAL1$^{*}$ promoter is carried out by both the activators (GAL) and
repressors (TetR). They have identified a regulatory mechanism and
key reactions using stochastic simulations that agree well with their
experimental observations \cite{key-03}. The important property of
their regulatory mechanism is that the activators and repressors can
bind the promoter simultaneously and non-competitively. Their observations
also revealed that the pulsatile mRNA production through the reinitiation
of transcription by RNAP II is crucial to match the experimental data
points of the Fano factor\footnote{The Fano factor is a measure of noise. The Fano factor and noise strength
are synonymous throughout the paper. For more, refer to glossary.} at the protein levels. Sanchez \textit{et al.} \cite{key-13} reproduced
the experimental results of \cite{key-03} by exact analytical calculation
in which the reinitiation of transcription process is mapped by average
burst distribution.

In this article, we consider the more general regulatory architecture
regulated by activator-repressor with non-competitive interaction
with the gene along with the reinitiation dynamics. It is noteworthy
that similar network was studied experimentally by \cite{key-03,key-06}.
In this work, we do study the same (four-state) network although our
approach is completely analytical. We find the exact analytical expressions
for mean and the Fano factor of mRNAs and proteins. These analytical
expressions are important to find the behaviors of mean and the Fano
factor with different rate constants and regulatory parameters.

The availability of exact analytical expression for any experimentally
measurable quantity is crucial in identifying the structure and function
of the complex cellular system. The average expression level \cite{key-48},
and the Fano factor \cite{key-35,key-50,key-51} are the important
physical quantities to identify the functional role of a complex gene
regulatory network. The exact analytical expressions of these biologically
significant quantities in terms of the rate constants of the biochemical
reactions of the network are, therefore, powerful tools for research. 

Here we study the transcriptional regulatory networks with non-competitive
architecture and analytically calculate the mean and Fano factor of
mRNAs and proteins for the network with and without the reinitiation
of transcription by RNAP II. The theory enables us to study the characteristics
of the aforementioned quantities with any of the parameters individually.
Thus, we search the dependence of the Fano factor (at mRNA level)
on reaction rate constants and find the Fano factor in the sub-Poissonian
regime by means of transcription reinitiation. We also reveal some
other factors that control the mean and noise for the network. Additionally,
we illustrate the effect of transcriptional reinitiation along with
other factors, specifically, the role of aTc on the network. By using
the analytical theory and simulation we are able to reproduce the
curves of mean and the noise that were previously found in \cite{key-03}.
We also notice that there is a mismatch of experimental data with
the theoretical curve proposed by Blake \textit{et al.} \cite{key-03}
(see figure \ref{fig:Non-Com-mean}c). We consider some extra transitions
to match the theory properly with the experimental results. With the
help of our analytical computation, we are able to find out the probable
set of rate constants that gives a good fit of experimental points
to the theoretical curves. We also perform statistical error minimization
technique and determine the sensitivity and uncertainties of the parameters
that establish the robustness of our analysis. Finally, we observe
some anomalies in noise curves of mRNA and in the variance of protein
at low activator (GAL) concentrations in presence of the repressor
molecule bounded by aTc.

\section{Non-competitive regulatory architecture and its analysis}

Regulation of transcription by activator and repressor is a well known
mechanism of gene regulation in the cell \cite{key-03,key-06,key-18,key-29,key-31}.
There are experimental evidences that transcriptional regulation by
activator and repressor can occur either non-competitively or competitively
\cite{key-03,key-10}. Blake \textit{et al.} \cite{key-03} studied
the synthetic yeast GAL1$^{*}$ promoter experimentally and observed
the variation of the Fano factor with respect to transcriptional efficiency,
defined as the ratio of transcription to the maximum transcription
\cite{key-13}. They also identified the architecture of transcriptional
regulatory network for the synthetic GAL1$^{*}$ promoter of yeast
by stochastic simulation.

\begin{figure}[H]
\begin{centering}
\includegraphics[width=6.5cm,height=4.5cm]{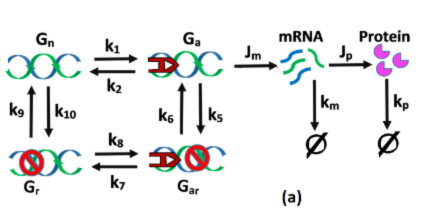}\qquad{}\includegraphics[width=7cm,height=5cm]{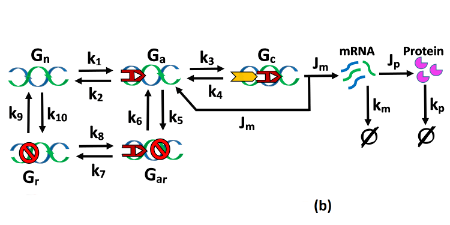}
\par\end{centering}
\vspace{1bp}

\begin{centering}
\includegraphics[width=7cm,height=5cm]{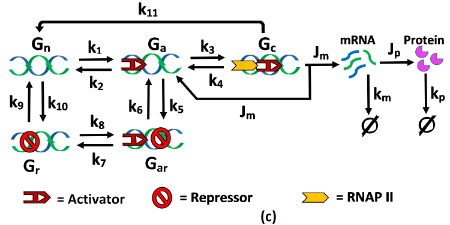}
\par\end{centering}
\caption{\label{fig:Non-competetive network}Reaction schemes for an activator-repressor
system with non-competitive interaction of a gene. (a) Network without
transcription reinitiation path: an activator binds to the promoter
of gene at normal state ($G_{n})$ to make the gene active ($G_{a})$
for synthesis of mRNA. Proteins are then produced from mRNA via reaction
rate $J_{p}.$ $k_{m}$ and $k_{p}$ are the rates of decay of mRNA
and proteins respectively. A repressor can bind to the normal state
to make a repressed state $(G_{r})$ which is unable to transcribe.
A repressor may also attach with an active state to bring the gene
at an active-repressed state ($G_{ar}).$\protect \\
 (b) Network with a transcription reinitiation path from an initiation
complex $G_{c}$ to $G_{a}$ via $J_{m}$. The state $G_{c}$ is formed
when RNAP II binds to an active state ($G_{a}).$ \protect \\
 (c) With transcription reinitiation path and a direct transition
path from the closed complex $G_{c}$ to the normal state $G_{n}$.\textcolor{orange}{{} }}
\end{figure}

The important property of the constructed promoter is that both the
activators (GAL) and repressors (TetR) interact with the gene non-competitively.
So, there can be four different states of the gene namely, normal
($G_{n}$), active ($G_{a}$), active-repressed ($G_{ar}$) and repressed
($G_{r}$) (Figure \ref{fig:Non-competetive network}a). The normal
state is the open or vacant state of promoter where either activator
or repressor can bind non-competitively. If the activator (repressor)
binds first then the normal state turns into an active (repressed)
state. A repressor (activator) can bind the active (repressed) state
of the promoter and turns it into an active-repressed state. At this
stage, anhydrotetracycline (aTc) binds with tetR to inhibit expression
and produces noise in the expression. This repression is modeled by
the aTc-dependent transition rate $k_{10}\propto\frac{(tetR)^{2}}{[1+(C_{i}*aTc)^{4}]^{2}}$
from $G_{n}$ to $G_{r}$ \cite{key-06} and $k_{5}=ek_{10}$ from
$G_{a}$ to $G_{ar}$ \cite{key-03}\footnote{Where $C_{i}$ is a constant and the factor \textit{e} appearing here
eventually controls the transitions from $G_{a}$ to $G_{ar}$ via
$k_{5}$ and from $G_{r}$ to $G_{ar}$ via $k_{8}=ek_{1}.$ Hence,
it affects the noise strength. We will explore it later in section
(2.4).} . At low aTc concentration, GAL1{*} promoter resides with high probability
in a repressed state with TetR bound resulting in low protein levels
and low levels of noise in protein production. In contrast, at high
inducer concentration, TetR is rarely bound and transcription is frequently
initiated resulting in high expression levels and less noise in protein
output. However, at intermediate levels of induction, the promoter
is more likely to transition between an active state and a repressed
state. A stable transcription form increases the possibility that,
once in the active state, the promoter will remain active, repeatedly
recruiting RNA Pol II in the course of reinitiation of transcription
\cite{key-24,key-34} and production of new transcripts.

Along with the four different states of the promoter, Blake \textit{et
al.} \cite{key-03} also identified that reinitiation of transcription
by RNAP II from the active state of the promoter is crucial to reproduce
the experimental data with stochastic simulation results \cite{key-03}.
The RNAP II binds the activated ($G_{a}$) gene and forms an initiation
complex ($G_{c}$). Then RNAP II starts transcription along the gene
and the close-complex turns into an activated state where another
RNAP II can bind. The reinitiation of transcription by RNAP II is
shown in figure \ref{fig:Non-competetive network}(b).

For the synthetic yeast GAL1$^{*}$ promoter, some rate constants
for similar transitions among the promoter states are assumed to be
correlated with each other. To make our study more general, we assume
that all the rate constants are completely uncorrelated with each
other (figure \ref{fig:Non-competetive network}). We have also incorporated
the possibility of direct transition from the close-complex ($G_{c}$)
to the normal state ($G_{n}$) \cite{key-27} as shown in figure \ref{fig:Non-competetive network}(c).
The introduction of that transition path is due to the fact that,
both activator and RNAP II can dissociate simultaneously from $G_{c}$
and bring back the gene to its normal state ($G_{n}).$

It should be stressed that, the reaction scheme of the gene regulatory
network in figure \ref{fig:Non-competetive network}(c) is the more
general one compared to figure \ref{fig:Non-competetive network}(b).
So, we write the Master equation \cite{key-32} corresponding to the
reaction scheme in figure \ref{fig:Non-competetive network}(c) and
calculate the mean and the Fano factor of mRNAs and proteins at the
steady state.

We assume that there exists $l$ copy number of a particular gene
in the cell. Let us consider $p(n_{1},n_{2},n_{3},n_{4},n_{5},n_{6},t)$
be the probability that at time $t$, there are $n_{5}$ number of
mRNAs and $n_{6}$ number of protein  molecules with $n_{1}$ number
of genes in the active state ($G_{a}$), $n_{2}$ number of genes
in the initiation complex ($G_{c}$), $n_{3}$ number of genes in
the active-repressed state ($G_{ar}$) and $n_{4}$ number of genes
in the repressed ($G_{r}$) state. The number of gene in the normal
states ($G_{n}$) are $(l-n_{1}-n_{2}-n_{3}-n_{4})$. The time evaluation
of the probability is given by

\begin{equation}
\begin{array}{ccc}
\frac{\partial p(n_{i},t)}{\partial t} & = & k_{1}[\{l-(n_{1}-1+n_{2}+n_{3}+n_{4})\}p(n_{1}-1,n_{2},n_{3},n_{4},n_{5},n_{6},t)\\
 &  & -\{l-(n_{1}+n_{2}+n_{3}+n_{4})\}p(n_{i},t)]\\
 &  & +k_{2}[(n_{1}+1)p(n_{1}+1,n_{2},n_{3},n_{4},n_{5},n_{6},t)-n_{1}p(n_{i},t)]\\
 &  & +k_{3}[(n_{1}+1)p(n_{1}+1,n_{2}-1,n_{3},n_{4},n_{5},n_{6},t)-n_{1}p(n_{i},t)]\\
 &  & +k_{4}[(n_{2}+1)p(n_{1}-1,n_{2}+1,n_{3},n_{4},n_{5},n_{6},t)-n_{2}p(n_{i},t)]\\
 &  & +k_{5}[(n_{1}+1)p(n_{1}+1,n_{2},n_{3}-1,n_{4},n_{5},n_{6},t)-n_{1}p(n_{i},t)]\\
 &  & +k_{6}[(n_{3}+1)p(n_{1}-1,n_{2},n_{3}+1,n_{4},n_{5},n_{6},t)-n_{3}p(n_{i},t)]\\
 &  & +k_{7}[(n_{3}+1)p(n_{1},n_{2},n_{3}+1,n_{4}-1,n_{5},n_{6},t)-n_{3}p(n_{i},t)]\\
 &  & +k_{8}[(n_{4}+1)p(n_{1},n_{2},n_{3}-1,n_{4}+1,n_{5},n_{6},t)-n_{4}p(n_{i},t)]\\
 &  & +k_{9}[(n_{4}+1)p(n_{1},n_{2},n_{3},n_{4}+1,n_{5},n_{6},t)-n_{4}p(n_{i},t)]\\
 &  & +k_{10}[\{l-(n_{1}+n_{2}+n_{3}+n_{4}-1)\}p(n_{1},n_{2},n_{3},n_{4}-1,n_{5},n_{6},t)\\
 &  & -\{l-(n_{1}+n_{2}+n_{3}+n_{4})\}p(n_{i},t)]\\
 &  & +k_{11}[(n_{2}+1)p(n_{1},n_{2}+1,n_{3},n_{4},n_{5},n_{6},t)-n_{2}p(n_{i},t)]\\
 &  & +J_{m}[(n_{2}+1)p(n_{1}-1,n_{2}+1,n_{3},n_{4},n_{5}-1,n_{6},t)-n_{2}p(n_{i},t)]\\
 &  & +k_{m}[(n_{5}+1)p(n_{1},n_{2},n_{3},n_{4},n_{5}+1,n_{6},t)-n_{5}p(n_{i},t)]\\
 &  & +J_{p}[n_{5}p(n_{1},n_{2},n_{3},n_{4},n_{5},n_{6}-1,t)-n_{5}p(n_{i},t)]\\
 &  & +k_{p}[(n_{6}+1)p(n_{1},n_{2},n_{3},n_{4},n_{5},n_{6}+1,t)-n_{6}p(n_{i},t)]
\end{array}\label{eq:NC-WR-1}
\end{equation}

where, i = 1,2,......, 6

We can derive the mean, variance and the Fano factor of mRNAs and
proteins from the moments of equation (\ref{eq:NC-WR-1}) with the
help of a generating function\footnote{refer to Appendix-A}. The
mean mRNA and protein are given by

\begin{equation}
m^{WR}=\frac{J_{m}k_{3}k_{8}b_{9}b_{2}}{k_{m}(-k_{3}k_{8}(b_{11}b_{1}+b_{10}k_{10})-(b_{11}b_{13}-b_{10}b_{12})b_{2})};\quad p^{NCWR}=\frac{m^{WR}\,J_{p}}{k_{p}}\label{eq:NonWR-M-2}
\end{equation}

\begin{equation}
FF_{m}^{WR}=1+A-m^{WR}\label{eq:NC-FFm-3}
\end{equation}

\begin{equation}
FF_{p}^{WR}=1+B-p^{WR}\label{eq:NC-FFp-4}
\end{equation}

where the detail expressions of A, B and C along with other $b_{j}$(j=1,2,......,20)
parameters are given in Appendix-B.

In the following analyses we have emphasized only on a subset of the
set of parameters that appear in equation (\ref{eq:NC-WR-1}). In
this way, we have successfully established how the same set of parameters
considered in \cite{key-03} also affects the physical quantities
that we can calculate for the configuration \ref{fig:Non-competetive network}(c).
An additional advantage in working with the same set of parameters
is that, in the limiting case we can reproduce the results of \cite{key-03,key-26,key-28}.

$\vphantom{}$

\begin{figure}[H]
\begin{centering}
\includegraphics[width=5.5cm,height=3.5cm]{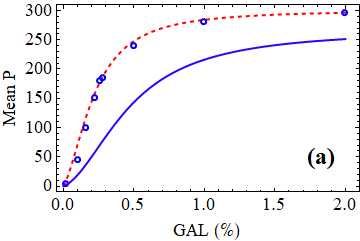} \includegraphics[width=5.5cm,height=3.5cm]{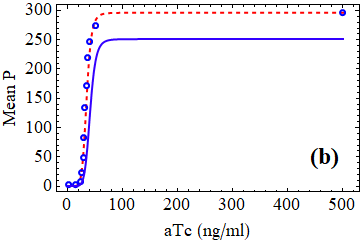}
\par\end{centering}
\begin{centering}
$\vphantom{}$
\par\end{centering}
\begin{centering}
\includegraphics[width=5.5cm,height=3.5cm]{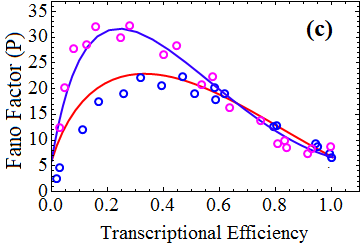}\includegraphics[width=5.5cm,height=3.5cm]{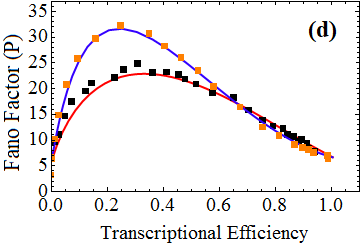}\includegraphics[width=5.5cm,height=3.5cm]{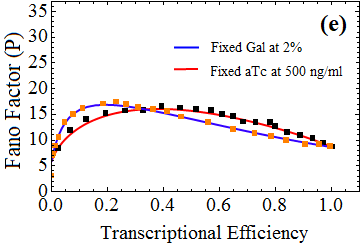}
\par\end{centering}
\caption{\label{fig:Non-Com-mean}Variation of mean protein with (a) GAL at
full aTc and (b) aTc at 2\% GAL. Solid (dashed) line is drawn from
analytical calculation corresponding to scheme \ref{fig:Non-competetive network}(a)
(figure \ref{fig:Non-competetive network}(b)). Hollow circles are
the experimental data points with 2\% GAL concentration in (b) and
with aTc=500 ng/ml in (a).\protect \\
 (c) and (d) represent the variation of the Fano factor with transcriptional
efficiency. In (c), blue (red) solid line is drawn analytically with
2\% GAL concentration (with aTc=500 ng/ml) from the scheme \ref{fig:Non-competetive network}(b).
In (d), blue (red) solid line is drawn analytically with 2\% GAL concentration
(with aTc=500 ng/ml) from configuration \ref{fig:Non-competetive network}(b).
The black and brown squares are generated from stochastic simulation
using the Gillespie algorithm \cite{key-33} from the reactions in
figure \ref{fig:Non-competetive network}(c). In (e), the red (blue)
solid line is from exact analytical expression and squares are from
stochastic simulation according to the configuration \ref{fig:Non-competetive network}(a). }
\end{figure}

In Figure \ref{fig:Non-Com-mean}, we have shown the variation of
mean and the Fano factor of proteins with respect to external inducers
and transcriptional efficiency. In the next step, we verify our analytical
results by matching the expressions of mean and the Fano factor calculated
from the reaction schemes in figure \ref{fig:Non-competetive network}
with the experimental data points in \cite{key-03}. Note that, in
this analysis we have considered the rate constants and the reaction
scheme of \cite{key-03} which corresponds to the figure \ref{fig:Non-competetive network}(b).
We also draw the curves for the mean (\ref{fig:Non-Com-mean}(a,b))
and the Fano factor (\ref{fig:Non-Com-mean}e) without the reinitiation
process\footnote{refer to Appendix-C for the corresponding analytical expressions },
correspondig to configuration \ref{fig:Non-competetive network}(a).
Subsequently, we compare the effect of reinitiation of transcription
on those two quantities. We find similar conclusion as in \cite{key-03}
that reinitiation can increase both the noise and mean expression
level. Both the experimental data and our analytical results show
that the reinitiation of transcription process in the present transcriptional
regulatory architecture increases the Fano factor at the protein level
(figure \ref{fig:Non-Com-mean}(d) and (e)). We also see that the
reinitiation of transcription increases the mean protein level compared
to the transcription without reinitiation \cite{key-28}. The rate
constants used in \cite{key-03} are given by $k_{1}=0.02+0.2*GAL,$
$k_{2}=0.01+0.1*GAL+0.077/GAL$,\,$k_{3}=50,k_{4}=10,k_{5}=e*k_{10},\,k_{10}=200*(npt)^{2}/[1+(C_{i}*aTc)^{4}]^{2},\:k_{6}=k_{9}=10,$$\,k_{8}=e*k_{1},k_{7}=k_{2},J_{m}=1,k_{m}=1,J_{p}=5,k_{p}=0.0125,npt=100,C_{i}=0.1,e=0.025$.
A better fitting with a different set of rate constants (using reaction
scheme \ref{fig:Non-competetive network}c) is shown in the following
paragraph by considering extra intermediate possible transitions.
We have analytically find the idea of another set of rate constants
that can give a better fitting of data. 

In figure \ref{fig:Non-Com-mean}(c), we see that the analytical curve
for variation of the Fano factor with transcriptional efficiency for
fixed aTc at 500 ng ml$^{-1}$ differ greatly with experimental data
points at lower values of transcriptional efficiency. Now we tried
for a different set of rate constants to remove this discrepancy.
We use the reaction scheme shown in figure \ref{fig:Non-competetive network}(c)
where a direct path of possible transition from $G_{c}$ to $G_{n}$
via $k_{11}$ has been considered, as both the activator and RNAP
II can remove simultaneously from the stage $G_{c}$ to bring back
the gene at stage $G_{n}.$ Also, the intermediate genetic stages
between G$_{n}$ and $G_{c}$ are considered which were ignored earlier. 

We assume when the activator molecules are attached to $G_{n}$, there
exists an intermediate state $G_{s}$ (Gene-dox complex). The activation
rate constant $k_{A}$ carries dox {[}s{]} whereas deactivation rate
$k_{D}$ releases dox from $G_{s}.$ We choose $k_{A}=k_{a}[s]$ and
$k_{D}=k/k_{A}=k_{d}/[s]$, where $k$ is a constant of proportionality.
There are a direct basal path $k_{B}$ (forward) and $k_{R}$ (reverse)
from $G_{n}$ to $G_{a}.$

Let us consider : 
\begin{figure}[H]
\centering{}\includegraphics[width=6cm,height=4cm]{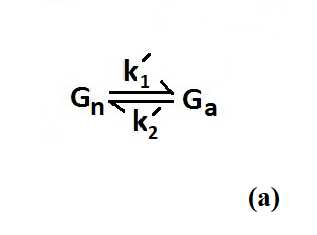} \hspace{4bp}\includegraphics[width=6cm,height=4cm]{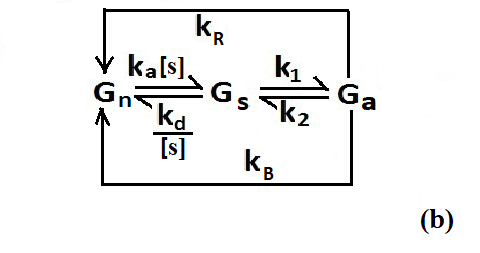}\caption{\label{fig:Intermediate-state-consideration}Intermediate state consideration:
(a) our consideration with equivalent reaction rates between normal
state $(G_{n})$ and active state ($G_{a})$. (b) generalized structure
with intermediate state $G_{s}$ (Gene-dox complex).}
\end{figure}

The kinetic equations are :

\begin{equation}
\frac{d[G_{a}]}{dt}=k_{1}[G_{s}]+k_{B}[G_{n}]-(k_{2}+k_{R})[G_{a}]\label{eq:inter state-1}
\end{equation}

\begin{equation}
\frac{d[G_{s}]}{dt}=k_{a}[s][G_{n}]-\frac{k_{d}}{[s]}[G_{s}]+k_{2}[G_{a}]-k_{1}[G_{s}]\label{eq:inter state-2}
\end{equation}

\begin{equation}
\frac{d[G_{n}]}{dt}=\frac{k_{d}}{[s]}[G_{s}]+k_{R}[G_{a}]-(k_{a}[s]+k_{B})[G_{n}]\label{eq:inter state-3}
\end{equation}

\begin{equation}
[G_{n}]+[G_{s}]+[G_{a}]=1\label{eq:inter state-4}
\end{equation}

Applying steady state condition, $\frac{d[G_{a}]}{dt}=0,\frac{d[G_{s}]}{dt}=0,\frac{d[G_{n}]}{dt}=0$
and solving we get

\begin{equation}
[G_{a}]=\frac{k_{1}^{'}}{k_{1}^{'}+k_{2}^{'}}\label{eq:inter state solution}
\end{equation}

where, 
\begin{equation}
k_{1}^{'}=k_{1}[k_{a}[s]+\frac{k_{B}}{k_{1}}\frac{k_{d}}{[s]}+k_{B}]\label{eq:equivalent k1}
\end{equation}

\begin{equation}
k_{2}^{'}=k_{2}^{c}[k_{a}[s]+\frac{k_{d}}{[s]}+k_{c}\label{eq:equivalent k2}
\end{equation}

with $k_{2}^{c}=(k_{2}+k_{R})$ and $k_{c}=\frac{k_{1}k_{R}+k_{2}k_{B}}{k_{2}^{c}}$.

\begin{figure}[H]
\begin{centering}
\includegraphics[width=5.5cm,height=3.5cm]{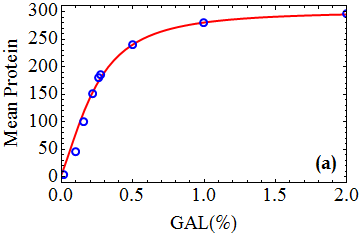} \includegraphics[width=5.5cm,height=3.5cm]{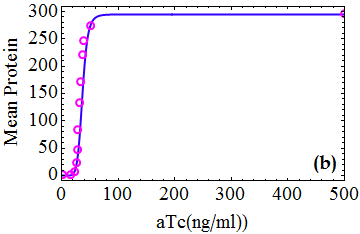}
\par\end{centering}
\begin{centering}
$\vphantom{}$
\par\end{centering}
\begin{centering}
\includegraphics[width=5.5cm,height=3.5cm]{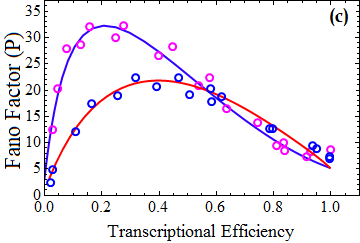} \includegraphics[width=5.5cm,height=3.5cm]{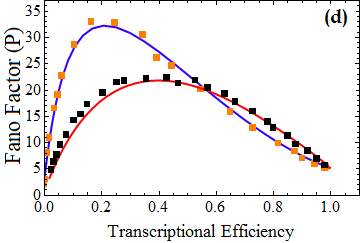}
\par\end{centering}
\caption{\label{fig:NC-with-K11}Variation of mean protein with (a) GAL at
aTc=500 ng/ml and (b) aTc at 2\% GAL. Solid lines are drawn from analytical
calculation corresponding to the figure \ref{fig:Non-competetive network}(c).
Hollow circles are the experimental data points with 2\% GAL concentration
in (b) and aTc=500 ng/ml in (a).\protect \\
 (c) and (d) shows the variation of the Fano factor with transcription
efficiency. Blue (red) solid line is drawn analytically with 2\% GAL
concentration (with aTc=500 ng/ml) from the configuration \ref{fig:Non-competetive network}(c).
In (d), the black and orange squares are are generated from stochastic
simulation using the Gillespie algorithm from the reactions in configuration
\ref{fig:Non-competetive network}(c) and the rate constants are given
in the text above.}
\end{figure}

When the intermediate state $G_{s}$ is absent, we have the same form
of $G_{a}$ as shown in equation (\ref{eq:inter state solution}).
Now from equations (\ref{eq:equivalent k1}) and (\ref{eq:equivalent k2})
we can see the exact form for the GAL dependent rate constants $k_{1}$
and $k_{2}$. Using trial and error method with different numerical
values we are able to fit the experimental data with theoretical curves.
The availability of the analytical expression of the Fano factor as
a function of different rate constants helps us to do that very easily.
The new set of rate constants are given by $k_{1}=0.0002/GAL+0.027+0.13*GAL,$
$k_{2}=0.002+0.1*GAL+0.06/GAL$, $k_{3}=50.0,k_{4}=12.5,k_{5}=e*k_{10},k_{10}=200*(npt)^{2}/[1+(C_{i}*aTc)^{4}]^{2},\:k_{4}=k_{9}=10,$$k_{11}=0.005$,$k_{7}=e*k_{1},k_{8}=k_{2},J_{m}=2.5,k_{m}=1,J_{p}=2.3,k_{p}=0.0125,npt=100,C_{i}=0.1,e=0.025$.
From figure (\ref{fig:NC-with-K11}) we observe that the analytical
curves for the variation of mean and the Fano factor well agree with
the experimental data points. With the new set of rate constants mentioned
above, the nature of variation of mean and the Fano factor at protein
level do not change. The best estimated values of parameters and their
uncertainties are discussed in Appendix-E.

\begin{figure}[H]
\begin{centering}
\includegraphics[width=6cm,height=4cm]{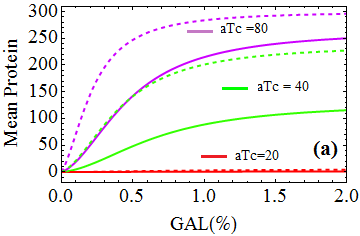} \includegraphics[width=6cm,height=4cm]{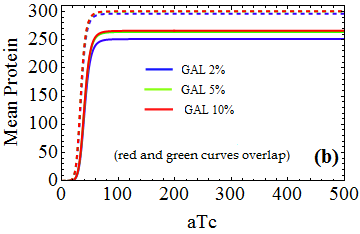} 
\par\end{centering}
\caption{\label{fig:Non-Com-Mean-diff-gal-atc}Variation of mean (a) with GAL
at different aTc concentration and (b) with aTc at different GAL (\%)
concentration. The solid curves correspond to the configuration \ref{fig:Non-competetive network}(a)
and the dashed curves are corresponding to the configuration \ref{fig:Non-competetive network}(b).}
\end{figure}

The plots figure (\ref{fig:Non-Com-Mean-diff-gal-atc}) show the variation
of mean protein with GAL (figure \ref{fig:Non-Com-Mean-diff-gal-atc}(a))
at different aTc and with aTc (figure \ref{fig:Non-Com-Mean-diff-gal-atc}(b))
at different GAL concentrations. It can be visualized that, for GAL
$\geq$2\% mean protein attains saturation at aTc \ensuremath{\approx}
60 $ng$ $ml^{-1},$ whereas for GAL < 2\% mean protein increases
rapidly for aTc > 20 $ng$ $ml^{-1}.$ Figure \ref{fig:Non-Com-Mean-diff-gal-atc}(b)
also showing that the mean protein is almost independent of GAL when
reinitiation is taken into consideration.

\begin{figure}[H]
\begin{centering}
\includegraphics[width=6cm,height=4cm]{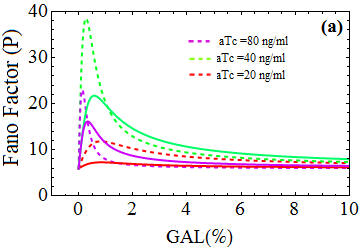} \includegraphics[width=6cm,height=4cm]{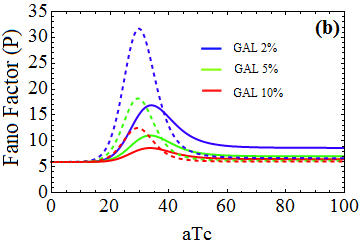}
\par\end{centering}
\caption{\label{fig:Non-Com-NS-aTc}Variation of the Fano factor with (a) GAL
(\%) at different aTc and (b) aTc at different GAL (\%) concentration.The
dashed curves are drawn from analytical calculations corresponding
to the scheme \ref{fig:Non-competetive network}(b). }
\end{figure}

\begin{figure}[H]
\begin{centering}
\includegraphics[width=6cm,height=4cm]{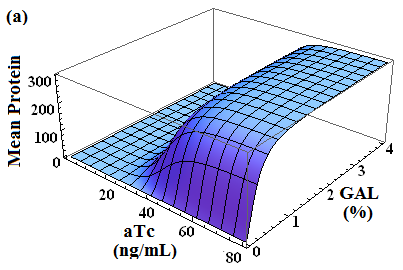} \includegraphics[width=6cm,height=4cm]{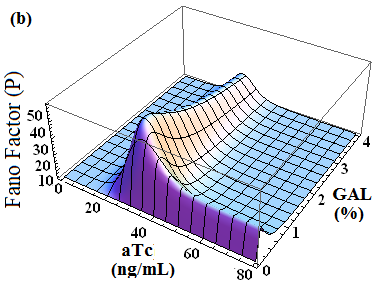}
\par\end{centering}
\begin{centering}
\includegraphics[width=6cm,height=4cm]{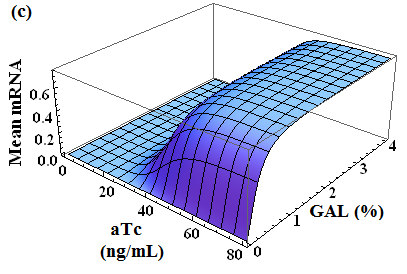} \includegraphics[width=6cm,height=4cm]{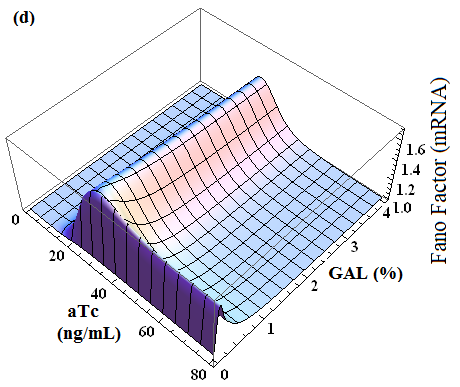}
\par\end{centering}
\caption{\label{fig:Non-comp-3D-Mean-NS}Variation of mean and the Fano factor
at protein and mRNA levels obtained from analytical calculations with
aTc and GAL corresponding to the scheme \ref{fig:Non-competetive network}(b).
The rate constants are chosen from Blake \textit{et al.} \cite{key-03}.
(a) and (b) are reproduced as in \cite{key-13}.}
\end{figure}

In figure \ref{fig:Non-Com-NS-aTc}(a), we plot the Fano factor with
GAL at different aTc. In figure \ref{fig:Non-Com-NS-aTc}(b), the
variations of the Fano factor with aTc for different GAL concentrations
are shown. It is clear from figure \ref{fig:Non-Com-NS-aTc}(a) that
the Fano factor is maximum at aTc = 40 ng ml$^{-1}$ as compared to
aTc = 20 and 80 ng ml$^{-1}$. The Fano factor is higher at lower
GAL concentration when observed with aTc variation. We present the
corresponding 3D plots in figure (\ref{fig:Non-comp-3D-Mean-NS})
to have a clearer view of the variation of mean and the Fano factor
of protein and mRNA with respect to GAL and aTc concentration. We
observe from figure (\ref{fig:Non-comp-3D-Mean-NS}) that both the
mean and the Fano factor vary with GAL and aTc. Howerver it is interesting
to note that, when we increase GAL concentration from 2\% to 10\%,
the Fano factor decreases accordingly (figure \ref{fig:Non-Com-NS-aTc}b).
On the other hand, figure \ref{fig:Non-Com-Mean-diff-gal-atc}(b)
shows that the changes in the mean protein level is extremely small
without reinitiation, whereas negligible changes are observed when
reinitiation of transcription is included and mean protein level becomes
nearly independent of GAL. Similar conclusion was drawn in \cite{key-13}
as well. However, here we have pointed out that this conclusion is
valid effectively only when reinitiation is involved (figure \ref{fig:Non-Com-Mean-diff-gal-atc}b
and \ref{fig:Non-Com-NS-aTc}b).

In figure \ref{fig:Non-comp-3D-Mean-NS}(a) and (b), we reproduce
the variation of mean protein and the Fano factor against aTc and
GAL (as in \cite{key-13}) by means of our analytical calculation.
Then we examine that the similar behavior of mean and the Fano factor
against GAL and aTc are followed by mRNA, as expected. We can see
there is a resonance type of incident for aTc \ensuremath{\approx}
30 $ng$ $ml^{-1}$ and GAL \ensuremath{\approx} 0.5\% (for the set
of rate constant given in \cite{key-03} which gives a sudden peak
in the Fano factor (noise strength)). The critical value of aTc, as
a function of GAL, for maximum Fano factor can be computed analytically.
The final expression is too large and hard to simplify, and hence
we avoid writing it here. Although, for a particular set of numerical
values of rate constants, we can determine the critical aTc value
from our analytical expression. 

\part*{{\normalsize{}2.1 The Fano factor in the sub-Poissonian region}}

The reinitiation process at the transcriptional level is crucial for
the reproduction of experimental results from stochastic simulation
of the model network in \cite{key-03}. Our analytical study of the
same model network shows that the reinitiation process increases the
Fano factor at the protein levels (figure \ref{fig:Non-Com-mean}e).
The effect of reinitiation is observed first at the mRNA levels. Figure
\ref{fig:Non-comp-3D-Mean-NS}(d) shows that the Fano factor at the
mRNA level also increases due to the reinitiation process for the
given rate constants. Our analysis also reveals that the reinitiation
process at the transcription level can bring down the Fano factor
at the mRNA level to the sub-Poissonian regime \cite{key-26,key-28}.
At this stage, we would like to check whether it is possible to observe
the sub-Poissonian Fano factor due to reinitiation dynamics in this
non-competitive regulatory architecture (configuration \ref{fig:Non-competetive network}b). 

In order to achieve that, we find the critical condition for $J_{m}$
by imposing the inequality $FF_{m}<1$ on equation (\ref{eq:NC-FFm-3}).
The resulting expression of critical $J_{m}$, i.e., $J_{m}^{c}$
is given by

\begin{equation}
J_{m}^{c}<\frac{B_{1}-B_{2}-B_{3}}{B_{4}-B_{5}+B_{6}}\label{eq:Jmc-5}
\end{equation}

\begin{flushleft}
where, 
\par\end{flushleft}

\begin{flushleft}
$B_{1}=b_{8}[b_{9}\{\left(b_{12}+k_{5}k_{m}\right)\left(k_{m}+k_{r}\right)-k_{3}k_{8}k_{10}\}$
\par\end{flushleft}

\hspace{20bp}$-k_{10}\left(b_{11}\left(b_{13}k_{r}+k_{3}\left(k_{4}-k_{1}\right)k_{8}\right)-b_{10}\left(b_{12}k_{r}-k_{3}k_{8}k_{10}\right)\right)]$
\begin{flushleft}
$B_{2}=\left(b_{16}\left(k_{m}+k_{1}\right)+b_{4}k_{1}k_{10}\right)\{b_{10}\left(b_{12}k_{r}-k_{3}k_{8}k_{10}\right)$
\par\end{flushleft}

$\hspace{20bp}-b_{11}\left(b_{13}k_{r}+k_{3}\left(k_{4}-k_{1}\right)k_{8}\right)\}$
\begin{flushleft}
$B_{3}=b_{9}[b_{13}b_{16}\left(k_{m}+k_{r}\right)-b_{16}k_{8}\left(k_{m}\left(k_{m}+k_{r}\right)+k_{1}k_{3}-k_{4}k_{3}\right)$
\par\end{flushleft}

$\hspace{20bp}+b_{4}k_{1}\left(b_{12}\left(k_{m}+k_{r}\right)+k_{5}k_{m}\left(k_{m}+k_{r}\right)-k_{3}k_{8}k_{10}\right)]$
\begin{flushleft}
$B_{4}=\left(b_{10}b_{12}-b_{11}\left(b_{13}+k_{3}k_{8}\right)\right)\left(b_{16}\left(k_{m}+k_{1}\right)+b_{4}k_{1}k_{10}\right)$
\par\end{flushleft}

\begin{flushleft}
$B_{5}=b_{8}\left(b_{9}\left(b_{12}+k_{5}k_{m}\right)+k_{10}\left(b_{10}b_{12}-b_{11}\left(b_{13}+k_{3}k_{8}\right)\right)\right)$
\par\end{flushleft}

\begin{flushleft}
$B_{6}=b_{9}\left(b_{16}k_{8}\left(k_{3}-k_{m}\right)+b_{4}k_{1}\left(b_{12}+k_{5}k_{m}\right)+b_{13}b_{16}\right)$
\par\end{flushleft}

and $k_{r}=k_{4}+k_{11}$. 

The details of $b_{j}$ parameters are given in Appendix-B.

We will see that these factors within $k_{r}$ have a big role in
controlling noise for the circuit.

\begin{figure}[H]
\begin{centering}
\includegraphics[width=6cm,height=4cm]{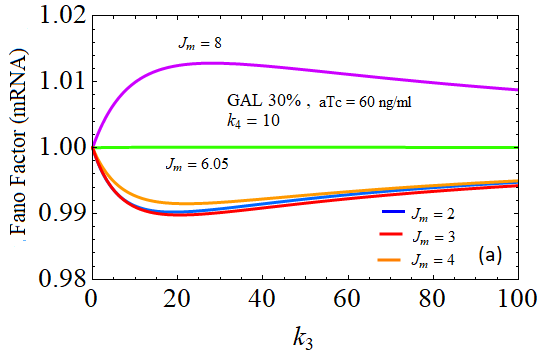} \includegraphics[width=6cm,height=4cm]{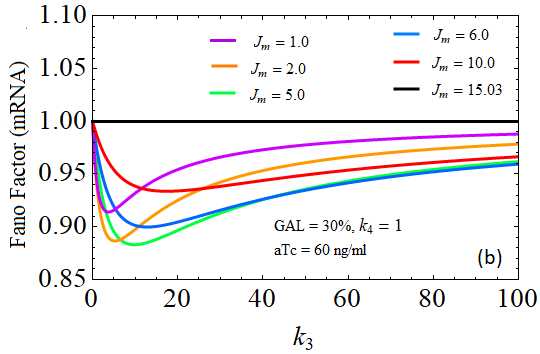}
\par\end{centering}
\caption{\label{fig:Critical Jm}Plot of Fano factor at mRNA level with $k_{3}$
for different values of $J_{m}$ with 30\% GAL concentrations and
aTc = 60 ng ml$^{-1}$. (a) The other rate constants are chosen from
Blake \textit{et al.} \cite{key-03}. (b) Here $k_{4}=1$. Other parameters
are the same as in \cite{key-03}. Lower value of $k_{4}$ helps to
reduce the Fano factor more below the Poissonian level.}
\end{figure}

\begin{figure}[H]
\begin{centering}
\includegraphics[width=6cm,height=4cm]{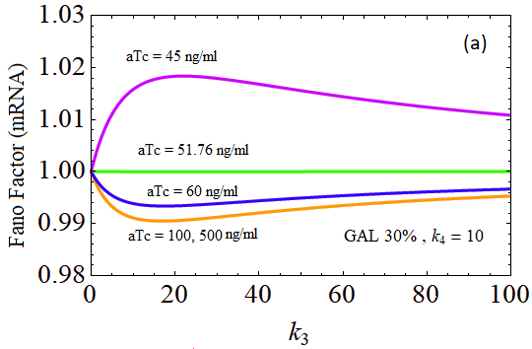} \includegraphics[width=6cm,height=4cm]{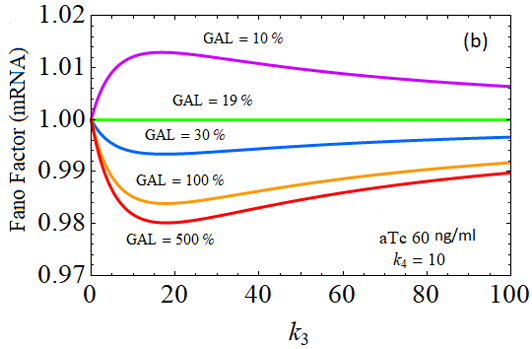}
\par\end{centering}
\caption{\label{fig:Critical_aTc_GAL}Plot of the Fano factor at mRNA level
with $k_{3}$ with (a) aTc and (b) GAL as parameter. The other rate
constants are chosen from Blake \textit{et al.} \cite{key-03}. }
\end{figure}

The presence of the critical value (which gives Poissonian Fano factor
at mRNA levels) $J_{m}^{c}$ in equation (\ref{eq:Jmc-5}) shows that
noncompetitive regulatory architecture with transcriptional reinitiation
can give rise to three different regimes of the Fano factor, viz.,
sub-Poissonian, Poissonian and super-Poissonian as shown in \cite{key-26,key-28}.
In figure (\ref{fig:Critical Jm}) we plot the three different regions
of the Fano factor along with the critical value of $J_{m}$ ($J_{m}^{c}$=
6.05) for the rate constants provided by Blake \textit{et al.} \cite{key-03}.
The plots show that the reduction of Fano factor towards the sub-Poissonian
regime is extremely small. However, the decrease in the value of the
Fano factor towards the sub-Poissonian regime is greater with lower
values of the rate constant $k_{4}$ (figure \ref{fig:Critical Jm}(b)).
We also find the critical values of aTc and GAL corresponding to the
rate constants in \cite{key-03}. Additionally, these three different
regimes of the Fano factor can also be observed for range of values
of aTc and GAL as shown in figure (\ref{fig:Critical_aTc_GAL}).

\begin{figure}[H]
\begin{centering}
\includegraphics[width=6cm,height=4cm]{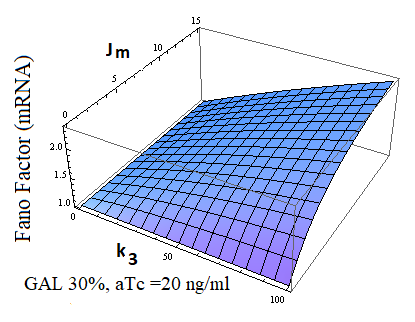} \includegraphics[width=6cm,height=4cm]{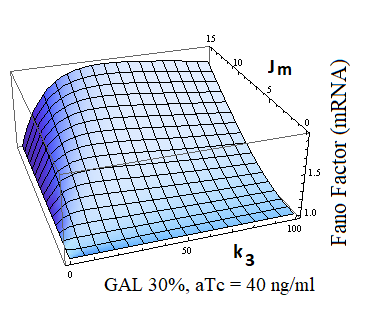}
\par\end{centering}
\begin{centering}
\includegraphics[width=6cm,height=4cm]{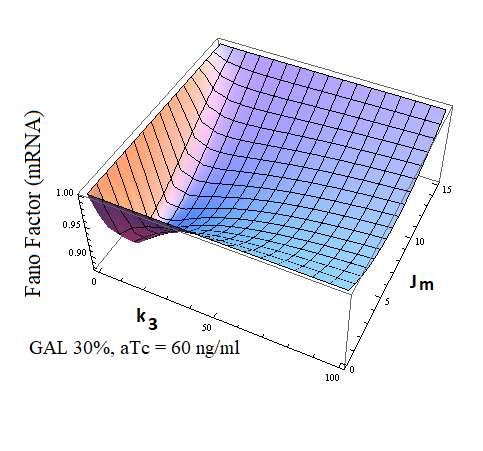} \includegraphics[width=6cm,height=4cm]{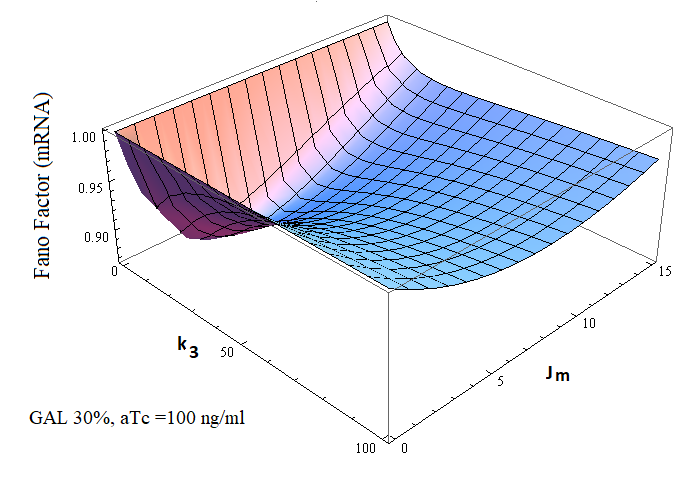}
\par\end{centering}
\caption{\label{fig:FFm_30_percent_GAL_diff_aTc}Variation of the Fano factor
at mRNA level with $k_{3}$ and $J_{m}$ for different aTc concentrations
with 30\% GAL and $k_{4}=1$.}
\end{figure}

\begin{figure}[H]
\begin{centering}
\includegraphics[width=6cm,height=4cm]{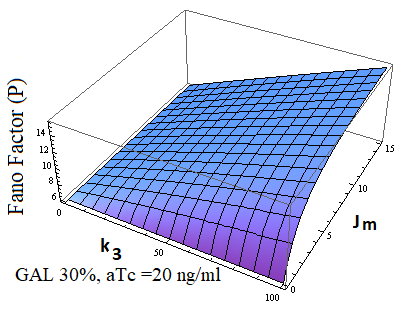} \includegraphics[width=6cm,height=4cm]{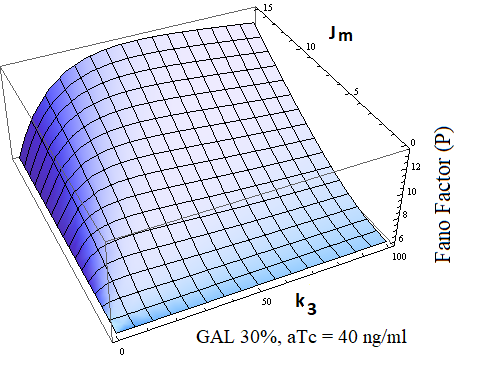}
\par\end{centering}
\centering{}\includegraphics[width=6cm,height=4cm]{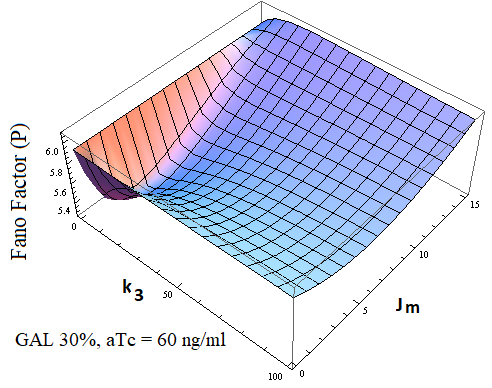} \includegraphics[width=6cm,height=4cm]{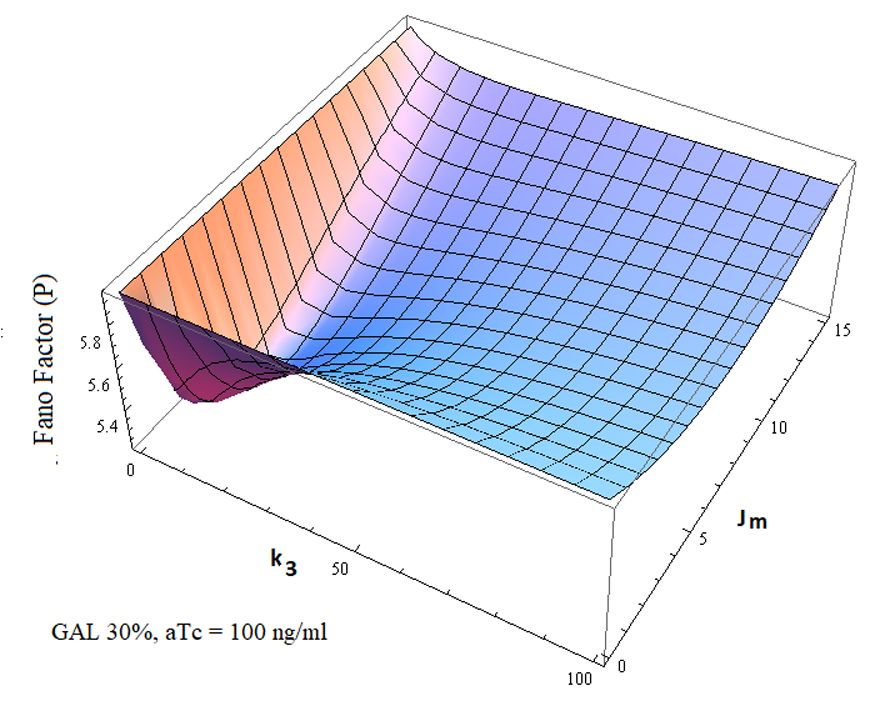}\caption{\label{FFp_30_percent_GAL_diff_aTc}Variation of the Fano factor at
protein level with $k_{3}$ and $J_{m}$ for different aTc concentrations
and fixed GAL concentration (30\%) and $k_{4}=1$. Here we find a
dip rather than the peak (as in Figure \ref{fig:Non-comp-3D-Mean-NS})
in the Fano factor at proten level due to reinitiation. }
\end{figure}

In order to observe the variation of the Fano factor more clearly,
we plot (3D) the Fano factor with $k_{3}$ and $J_{m}$ for different
concentrations of aTc and GAL. Figure \ref{fig:FFm_30_percent_GAL_diff_aTc}
show that the Fano factor can go below unity when aTc is more than
40 ng/ml with $30\%$ GAL concentrations. Consequently, we observe
a valley/dip in the Fano factor (at protein levels) at aTc more than
40 ng/ml and $30\%$ GAL concentration, in figure (\ref{FFp_30_percent_GAL_diff_aTc}).
This observation is in sharp contrast to figure (\ref{fig:Non-comp-3D-Mean-NS})
where we observe peaks in the Fano factor rather than dips. We have
also examined that the Fano factor at sub-Poissonian region as shown
in figure (\ref{fig:Critical_aTc_GAL}) can not be reduced below the
level presented by a two-state network with reinitiation \cite{key-26,key-28}.
Whatever may be the values of other rate constants at aTc$\geq$100,
our non-competitive model effectively reduces to a two-state model
and can reproduce all the features as shown in \cite{key-28}. This
can be analytically shown (refer to section 2.2) that our proposed
non-competitive architecture can be reduced to a two-state network
at aTc\textrightarrow \ensuremath{\infty} and $k_{11}=0.$

\part*{{\normalsize{}2.2 Role of aTc}}

Tetracycline (Tc) controlled gene expression has been exhibited in
a variety of eukaryotic systems including \textit{Saccaromyces cerevisiae
}\cite{key-57}\textit{.} Although Tc has some good medicinal properties,
its very little but distinct cytotoxicity to mammalian cells reduces
its applicability. Rather, one of its derivatives, anhydrotetracycline
(aTc), which binds the Tet repressor (TetR) more effectively than
Tc, is widely used in activator-repressor systems. It has also a lower
antibiotic activity towards E. coli \cite{key-52}. A tetR (repressor)
attached with aTc inhibits expression and produces noise in the expression.
This repression is modeled by the aTc dependent transition rate $k_{10}\propto\frac{(tetR)^{2}}{[1+(C_{i}*aTc)^{4}]^{2}}$
and $k_{5}=ek_{10}$.

\begin{wrapfigure}[11]{r}{0.4\columnwidth}%
\includegraphics[width=8cm,height=6cm]{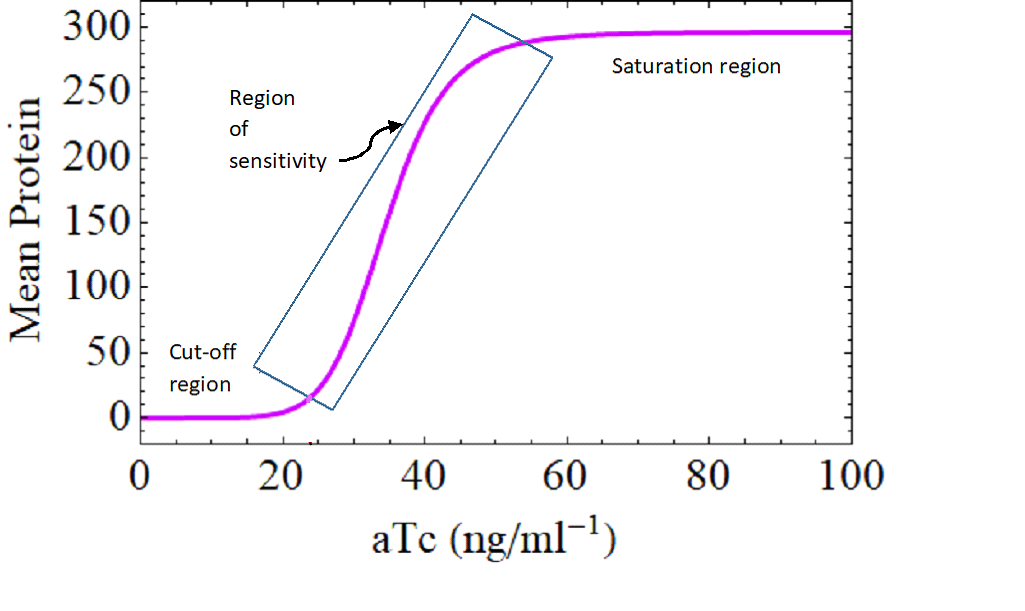}\caption{\label{fig:mean-expression-(protein)}Mean protein \textit{vs.} aTc }

\end{wrapfigure}%

\begin{table}[H]
\begin{centering}
\begin{tabular}{|c|c|c|c|c|c|c|c|c|c|}
\hline 
aTc (ng ml$^{-1})$ & 20 & 25 & 30 & 35 & 40 & 50 & 65 & 80 & 100\tabularnewline
\hline 
\hline 
$k_{10}$ & 6920 & 1250 & 297.53 & 87.64 & 30.28 & 5.01 & 0.78 & 0.119 & 0.019\tabularnewline
\hline 
\end{tabular}
\par\end{centering}
\caption{\label{tab:sensitivity-of-k_10}Sensitivity of $k_{10}$ to aTc concentration}

\end{table}

The response of the non-competitive circuit is very sensitive to aTc
concentration. Figure (\ref{fig:mean-expression-(protein)}) exhibits
a response (mean protein) versus aTc curve for non-competitive architecture.The
curve has sigmoid nature and looks very similar to a output characteristic
of a junction transistor having three region of operation \textit{viz.}
cut-off region, active region and saturation region. The active region
is the region of interest known as ``region of sensitivity''. In
cut-off region of operation ($0\leq aTc\leq25$) both the mean and
noise is negligible. In the active region ($(25\leq aTc\leq55$) mean
expression increases sharply offering a larger noise (figure \ref{fig:Non-Com-Mean-diff-gal-atc}b
and figure \ref{fig:Non-Com-NS-aTc}b). When the mean value reaches
to saturation (aTc>55) the noise reduces to a very low level. For
lower concentration of aTc the gene state is highly repressed that
resists to express. Figures (\ref{fig:Non-Com-Mean-diff-gal-atc})
to (\ref{FFp_30_percent_GAL_diff_aTc}) explains how aTc plays an
governing role in regulation of mean and noise. Table \ref{tab:sensitivity-of-k_10}
shows the sensitivity of $k_{10}$ to aTc concentration. A small change
in aTc results in a larger change in $k_{10}$. If we make aTc concentration
significantly high which effectively blocks the transitions : towards
$G_{r}$ from $G_{n}$ and towards $G_{ar}$ from $G_{a}$. For aTc\textrightarrow \ensuremath{\infty}
, both $k_{10}$ and $k_{5}$ (as $k_{5}=ek_{10})$ are effectively
zero. Then, along with $k_{11}=0$ the reaction scheme \ref{fig:Non-competetive network}(c)
reduces to a two-state model with reinitiation path involved.

Under these two limits we have from equation (\ref{eq:NonWR-M-2}),
\begin{equation}
m^{WR}=\frac{J_{m}k_{1}k_{3}}{k_{m}((J_{m}+k_{4})(k_{1}+k_{2})+(k_{1}k_{3}))}=\frac{J_{m}k_{1}k_{3}}{a_{2}^{'}k_{m}}\label{eq:16, T-S equi_meanM}
\end{equation}

and from equation (\ref{eq:NC-FFm-3}), 
\[
FF_{m}^{WR}=1+\frac{J_{m}k_{3}(k_{1}+k_{m})}{(a_{1}^{'}+a_{2}^{'})k_{m}}-\frac{J_{m}k_{1}k_{3}}{a_{2}^{'}k_{m}}
\]
\begin{equation}
=1+\frac{J_{m}k_{3}(a_{2}^{'}-a_{1}^{'}k_{1})}{(a_{1}^{'}k_{m}+a_{2}^{'})a_{2}^{'}}\label{eq:17, T-S equi FFm}
\end{equation}

Note that, equation (\ref{eq:16, T-S equi_meanM}) and equation (\ref{eq:17, T-S equi FFm})
have the same form of the mean mRNA and the Fano factor (mRNA) of
a two-state network with reinitiation given in \cite{key-26,key-28}
with $a_{1}^{'}=k_{1}+k_{2}+k_{3}+k_{4}+J_{m}+k_{m}$ and $a_{2}^{'}=(J_{m}+k_{4})(k_{1}+k_{2})+k_{1}k_{3}$.
Here we conclude that, practically, for aTc $\geq$100 and $k_{11}=0$
non-competitive network behaves like a two-state model producing all
the plots as shown in \cite{key-28}. 

\part*{{\normalsize{}2.3 Anomaly in the Fano factor (mRNA) at lower GAL
concentration: role of $\boldsymbol{k_{11}}$ }}

One of the major aspects of our work is the introduction of a direct
transition path from the stage $G_{c}$ to $G_{a}$ \cite{key-27},
to the model previously studied by Blake \textit{et al.} \cite{key-03,key-06}.
We subsequently study the behavior of mean expressions and noise (in
terms of the Fano factor) affected by the reaction rate $k_{11}$
for that path. As the gene expression is a complex process, we could
not deny the possibility of simultaneous unbound of RNAP II and activator
molecule and bring the gene to the normal state $G_{n}.$

\begin{figure}[H]
\includegraphics[width=5.5cm,height=3.5cm]{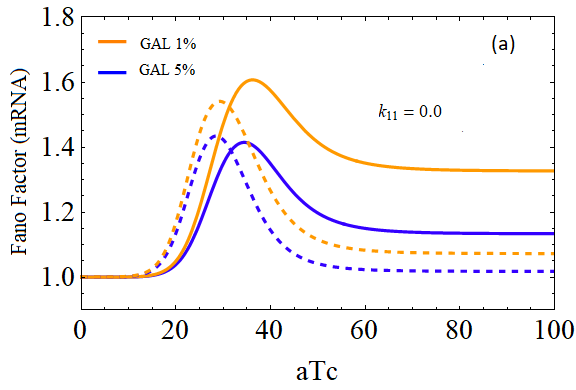} \includegraphics[width=5.5cm,height=3.5cm]{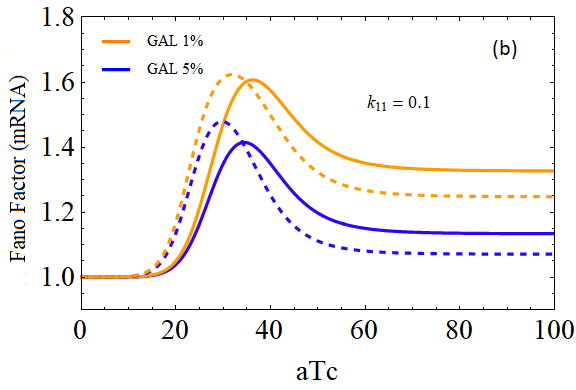}
\includegraphics[width=5.5cm,height=3.5cm]{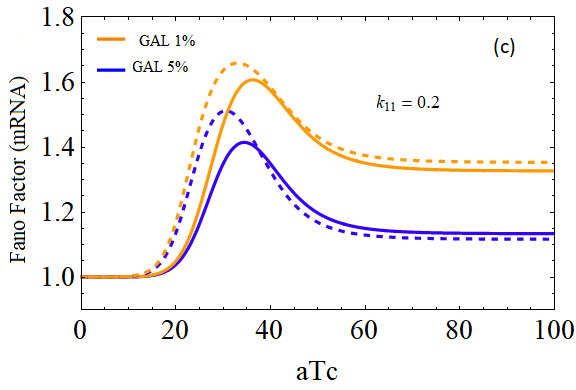}\caption{\label{fig:anomoly-of-noise}Variation of the Fano factor against
aTc for GAL 1\% and 5\% with different $k_{11}$, solid lines correspond
to configuration \ref{fig:Non-competetive network}(a) and dashed
lines correspond to configuration \ref{fig:Non-competetive network}(c).
Rate constants are chosen from \cite{key-03}.}
\end{figure}

With the rate constants used in \cite{key-03} along with $k_{11}=0$,
we notice from figure \ref{fig:anomoly-of-noise}(a) that, when GAL
< 5\%, the maximum value of the Fano factor (at mRNA level ) without
reinitiation is higher than that with reinitiation. The explanation
of this behavior is subject to further analysis which we have not
performed here. Nevertheless, we find that this anomaly goes off (i)
for higher GAL concentrations and (ii) for the introduction of small
non-zero value of $k_{11}$.

\noindent We further observe from figure \ref{fig:anomoly-of-noise}
that, the peak of the Fano factor curve (corresponding to the scheme
\ref{fig:Non-competetive network}c) has raised for $k_{11}=0.1$
and for $k_{11}=0.2$. This establishes the role of $k_{11}$ in raising
the Fano factor up a bit. A non-zero, positive value of $k_{11}>0$
reduces the effective transition probabilities via rate constants
$k_{2},k_{3},k_{4},k_{5},k_{6}$ and the transition from $G_{c}$
to mRNA via $J_{m}$ (see tables 3 and 4 in supplementary material).
That helps to raise the Fano factor and hence removes that anomalous
nature.

\begin{figure}[H]

\begin{centering}
\includegraphics[width=6cm,height=4cm]{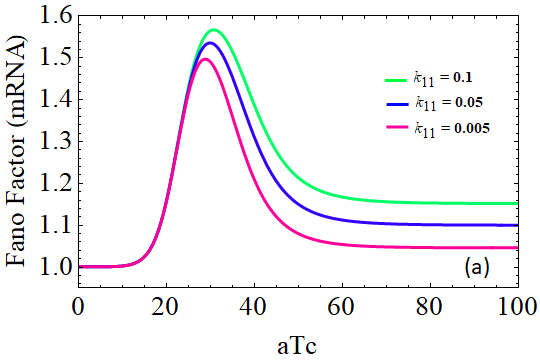} \includegraphics[width=6cm,height=4cm]{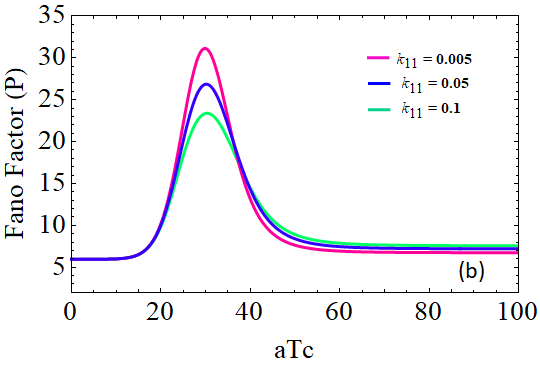}
\par\end{centering}
\begin{centering}
\includegraphics[width=6cm,height=4cm]{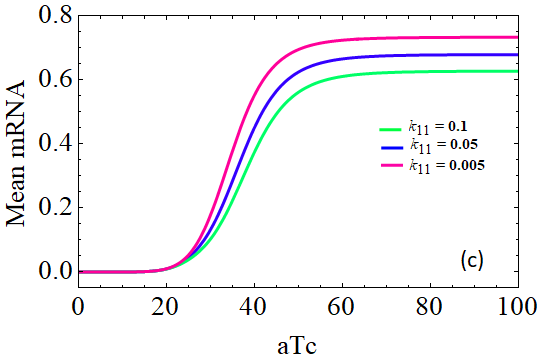} \includegraphics[width=6cm,height=4cm]{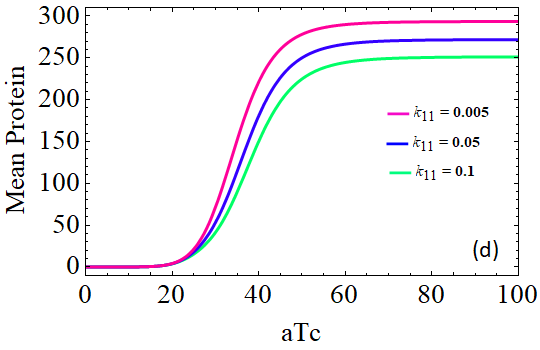}
\par\end{centering}
\caption{\label{fig:role-of-k11_2D}Role of $k_{11}$: (a) Fano factor (mRNA)
\textit{vs.} aTc, (b) Fano factor (protein) \textit{vs.} aTc, (c)
mean mRNA \textit{vs.} aTc, (d) mean protein \textit{vs.} aTc. \protect \\
In these plots, different $k_{11}$ values are taken as parameter
which reveal that, although mean values offering same behavior, the
Fano factors are showing slightly different nature in transcription
and translation level. Rate constants are chosen from \cite{key-03}.}
\end{figure}
\begin{figure}[H]
\begin{centering}
\includegraphics[width=6cm,height=4cm]{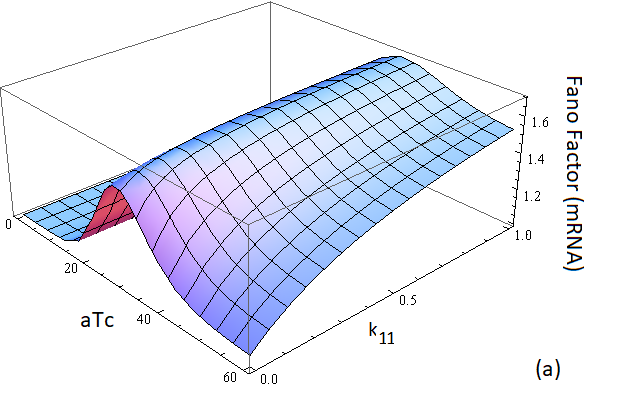} \includegraphics[width=6cm,height=4cm]{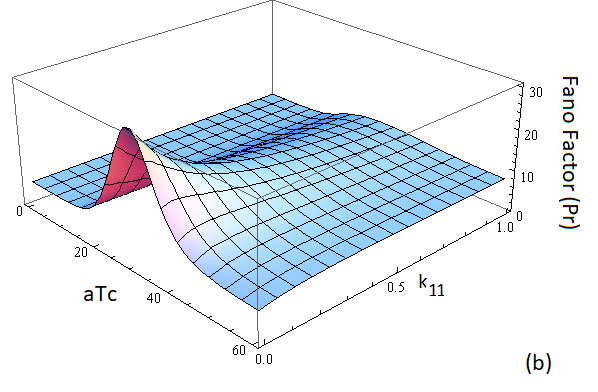}
\par\end{centering}
\begin{centering}
\includegraphics[width=6cm,height=4cm]{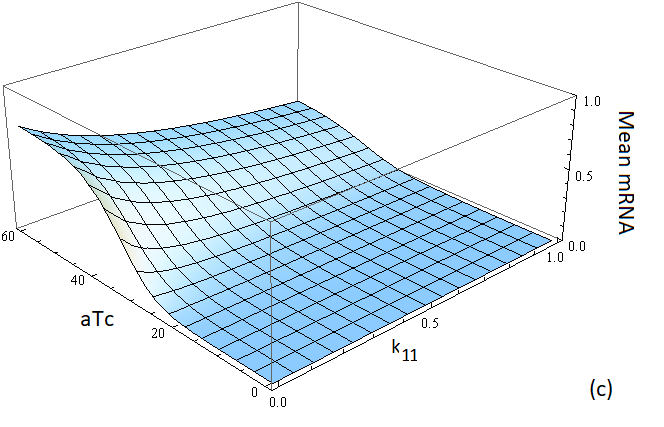} \includegraphics[width=6cm,height=4cm]{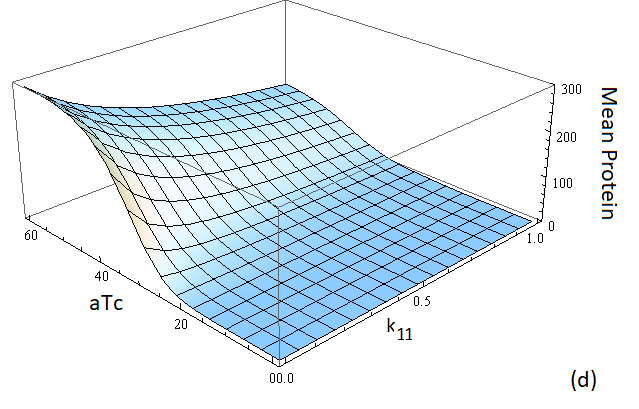}\caption{\label{fig:role-of-k11_3D}Role of $k_{11}$: 3D plot showing a clear
difference in Fano factors at mRNA and protein level while variations
of mean having similar nature. Rate constants are chosen from \cite{key-03}. }
\par\end{centering}
\end{figure}
There are another very interesting role of $k_{11}$ as shown in figure
(\ref{fig:role-of-k11_2D}) and figure (\ref{fig:role-of-k11_3D}).
The plots of mean expressions (mRNA and protein) against aTc are decreasing
with increasing $k_{11}$ values while the Fano factors at mRNA level
(transcriptional) and that at protein level (translational) show some
noticeable differences. The peaks of the Fano factors (mRNA level)
are higher for higher values of the parameter $k_{11}$ but the peaks
of the Fano factors at protein level decreases for higher values of
$k_{11}.$ On the other hand, the horizontal linear portion of the
Fano factor curve (right side of the curve \ref{fig:role-of-k11_2D}b
) almost merges for different $k_{11}$values (aTc > 40) for proteins
whereas the corresponding curves remain separated for mRNAs. This
implies that the noise is different for different $k_{11}$ values
at transcriptional level while it is almost same at translation level
at higher aTc values (>40 ng ml$^{-1})$.

\part*{{\normalsize{}2.4 Another noise reducing factor}}

\begin{figure}[H]
\centering{}\includegraphics[width=5.5cm,height=4cm]{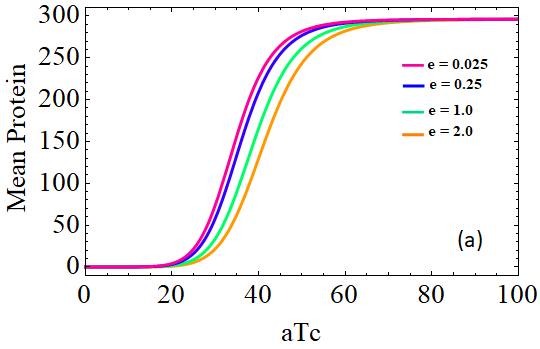} \includegraphics[width=5.5cm,height=4cm]{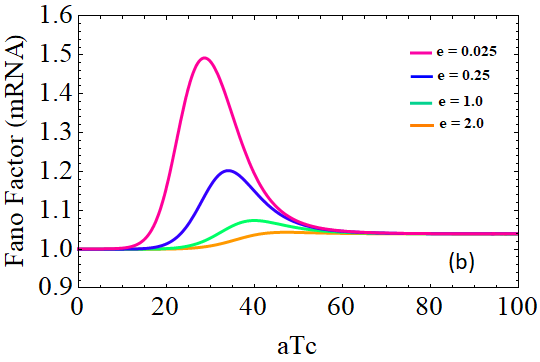}
\includegraphics[width=5.5cm,height=4cm]{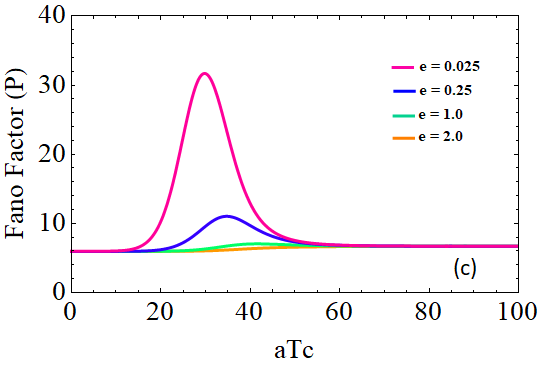}\caption{\label{fig:role-of-e}Role of the factor \textit{e} : (a) variation
of mean protein against aTc showing that saturation mean value does
not change with different values of \textit{e} except a little lateral
shift in the region of sensitivity. (b) The Fano factor (mRNA) against
aTc reduces with increasing value of \textit{e}. (c) The Fano factor
(protein) against aTc decreases with increasing \textit{e}. Other
rate constants are chosen from \cite{key-03}.}
\end{figure}

We have found that the noise, corresponding to the circuit \ref{fig:Non-competetive network}(c)
can be reduced with the help of the factor \textit{e} appearing in
the rate constants in \cite{key-03}. Without altering the maximum
of mean expression (figure \ref{fig:role-of-e}), we can reduce noise
(denoted by the Fano factor here) by changing the value of the factor
\textit{e}. Figure \ref{fig:role-of-e}(b) and (c) show that, an increasing
\textit{e} reduces the Fano factor much effectively in both transcription
and translation levels. It can be observed (figure \ref{fig:role-of-e}a)
that the maximum and also the saturation value of mean protein remains
unchanged with the increasing values of \textit{e} except a little
lateral shift in the ``region of sensitivity''. This also reveals
the fact that noise can be reduced when gene operates towards the
active-repressed compound state ($G_{ar}).$ Generally, an additional
genetic state (for linear circuits) causes much noise but here we
see that extra states (squared architecture) reduce noise.

\part*{{\normalsize{}2.5 Anomalous peak in the variance of Protein against
aTc }}

\begin{figure}[H]
\begin{centering}
\includegraphics[width=5.5cm,height=3.5cm]{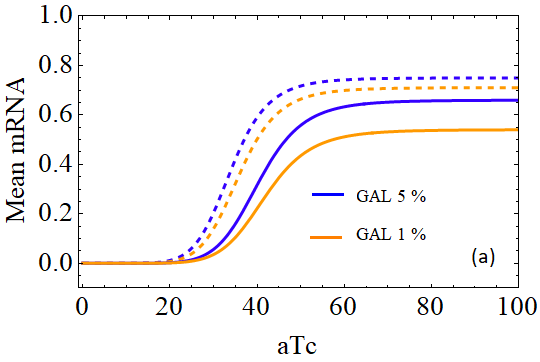} \includegraphics[width=5.5cm,height=3.5cm]{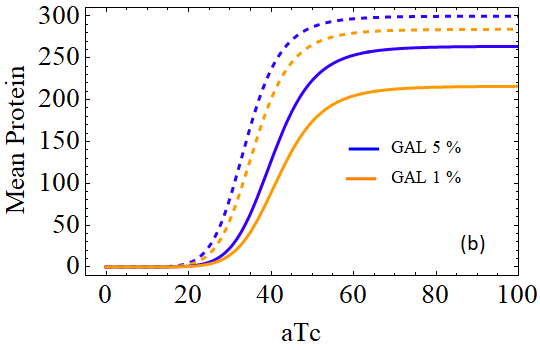}
\par\end{centering}
\begin{centering}
\includegraphics[width=5.5cm,height=3.5cm]{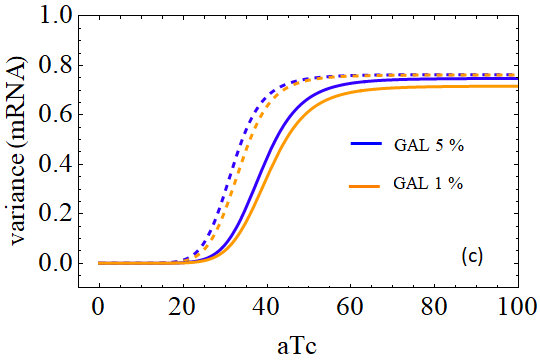} \includegraphics[width=5.5cm,height=3.5cm]{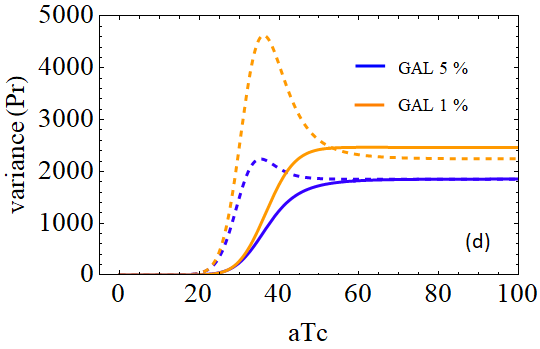}\caption{\label{fig:Variation-of-mean- and-variance}Variations of mean mRNA
and mean Protein against aTc are plotted in (a) and (b) respectively.
In (c) and (d) we plot of variance of mRNA and variance of protein
against aTc repectively. Solid lines correspond to cofiguration \ref{fig:Non-competetive network}(a)
and dashed lines correspond to configuration \ref{fig:Non-competetive network}(c).
Rate constants are chosen from \cite{key-03}.}
\par\end{centering}
\end{figure}

\begin{figure}[H]
\begin{centering}
\includegraphics[width=6cm,height=4cm]{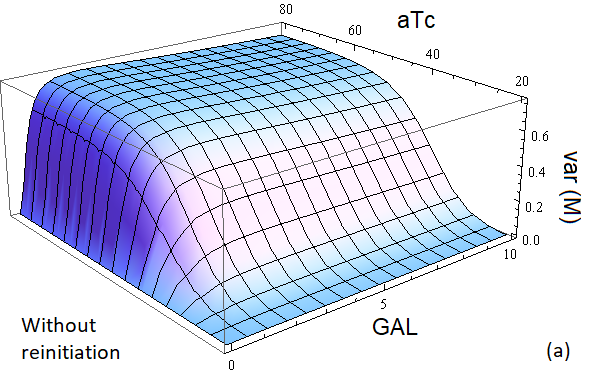} \includegraphics[width=6cm,height=4cm]{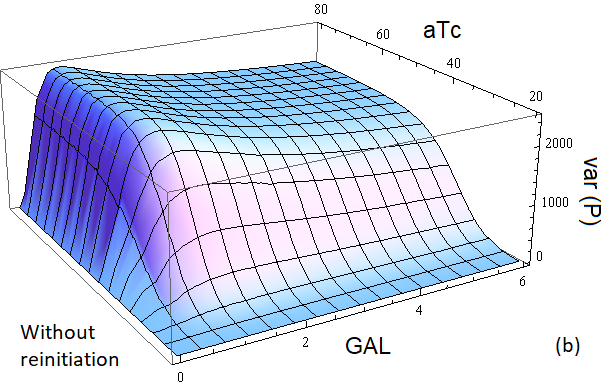}
\par\end{centering}
\centering{}\includegraphics[width=6cm,height=4cm]{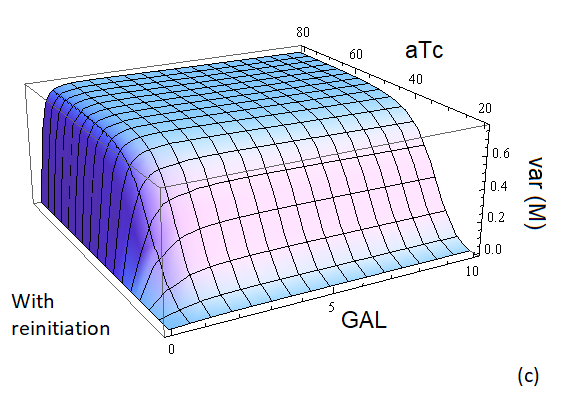} \includegraphics[width=6cm,height=4cm]{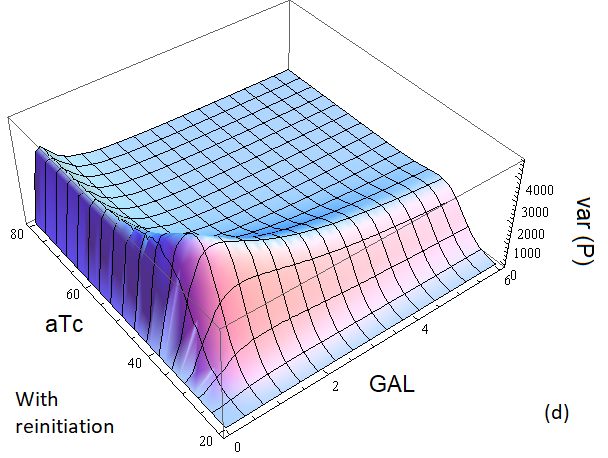}\caption{\label{fig:3D-variation-of -mean-variance}In (a) and (b) we plot
variance of mRNA (var (M)) and variance of protein (var (P)) with
GAL and aTc corresponding to configuration \ref{fig:Non-competetive network}(a)
and from the reaction scheme \ref{fig:Non-competetive network}(c)
we have the plots of variance at mRNA level (c) and that at protein
level (d) against GAL and aTc. Rate constants are chosen from \cite{key-03}.}
\end{figure}

On top of that, using the same rate constants as in \cite{key-03}
along with $k_{11}=0$ we also found a sudden peak in the variance
of protein against aTc (see figure \ref{fig:Variation-of-mean- and-variance}d)
in presence of trancriptional reinitiation corresponding to the process
\ref{fig:Non-competetive network}(b). This peak value increases with
the reduction in GAL concentration. We find it interesting that the
peak is not observed in variance of mRNA against aTc. Thus, one can
predict that the peak is due to the process of translation. But a
comparative 3D plot of variances of mRNA and protein (figure \ref{fig:3D-variation-of -mean-variance}),
with and without reinitiation (corresponding to the process \ref{fig:Non-competetive network}(a)
and \ref{fig:Non-competetive network}(b), respectively), shows that
the trancriptional reinitiation effect is responsible for it. We notice
that the peak is near aTc \ensuremath{\approx} 36.0 ng ml$^{-1}$.
We prepare the following table that contains a number of transitions
(for a single cell) between different genetic steps in gene expression
for different aTc values.

\begin{table}[H]

\begin{centering}
\begin{tabular}{|>{\centering}p{0.8cm}|>{\centering}p{0.8cm}|>{\centering}p{0.8cm}|>{\centering}p{0.8cm}|>{\centering}p{0.8cm}|>{\centering}p{0.8cm}|>{\centering}p{0.8cm}|>{\centering}p{0.8cm}|>{\centering}p{0.8cm}|>{\centering}p{0.8cm}|>{\centering}p{0.8cm}|>{\centering}p{0.8cm}|>{\centering}p{0.8cm}|}
\hline 
\multicolumn{13}{|c|}{(a) Without Reinitiation GAL 1\%}\tabularnewline
\hline 
aTc

{\footnotesize{}(ng ml$^{-1}$)} & {\small{}$G_{n}$to $G_{a}$}\\
{\small{}via $k_{1}$} & {\small{}$G_{a}$to $G_{n}$}\\
{\small{}via $k_{2}$} & {\small{}$G_{a}$to $G_{c}$}\\
{\small{}via $k_{3}$} & {\small{}$G_{c}$to $G_{a}$}\\
{\small{}via $k_{4}$} & {\small{}$G_{a}$to $G_{ar}$}\\
{\small{}via $k_{5}$} & {\small{}$G_{ar}$to $G_{a}$}\\
{\small{}via $k_{6}$} & {\small{}$G_{ar}$to $G_{r}$}\\
{\small{}via $k_{7}$} & {\small{}$G_{r}$to $G_{ar}$}\\
{\small{}via $k_{8}$} & {\small{}$G_{r}$to $G_{n}$}\\
{\small{}via $k_{9}$} & {\small{}$G_{n}$to $G_{r}$}\\
{\small{}via $k_{10}$} & {\small{}$G_{c}$to $M$}\\
{\small{}via $J_{m}$} & {\small{}$M$ to $P$}\\
{\small{}via $J_{p}$}\tabularnewline
\hline 
\hline 
{\footnotesize{}29.45} & {\footnotesize{}18} & {\footnotesize{}15} & {\footnotesize{}-----} & {\footnotesize{}-----} & {\footnotesize{}815} & {\footnotesize{}812} & {\footnotesize{}22} & {\footnotesize{}19} & {\footnotesize{}27319} & {\footnotesize{}27317} & {\footnotesize{}104} & {\footnotesize{}619}\tabularnewline
\hline 
{\footnotesize{}31.56} & {\footnotesize{}33} & {\footnotesize{}35} & {\footnotesize{}---} & {\footnotesize{}---} & {\footnotesize{}931} & {\footnotesize{}933} & {\footnotesize{}16} & {\footnotesize{}18} & {\footnotesize{}25885} & {\footnotesize{}25887} & {\footnotesize{}186} & {\footnotesize{}984}\tabularnewline
\hline 
{\footnotesize{}33.0} & {\footnotesize{}46} & {\footnotesize{}46} & {\footnotesize{}---} & {\footnotesize{}---} & {\footnotesize{}904} & {\footnotesize{}905} & {\footnotesize{}16} & {\footnotesize{}17} & {\footnotesize{}24612} & {\footnotesize{}24613} & {\footnotesize{}267} & {\footnotesize{}1410}\tabularnewline
\hline 
{\footnotesize{}34.73} & {\footnotesize{}60} & {\footnotesize{}63} & {\footnotesize{}---} & {\footnotesize{}---} & {\footnotesize{}684} & {\footnotesize{}684} & {\footnotesize{}12} & {\footnotesize{}15} & {\footnotesize{}23660} & {\footnotesize{}23664} & {\footnotesize{}330} & {\footnotesize{}1496}\tabularnewline
\hline 
{\footnotesize{}36.0} & {\footnotesize{}71} & {\footnotesize{}77} & {\footnotesize{}---} & {\footnotesize{}---} & {\footnotesize{}659} & {\footnotesize{}665} & {\footnotesize{}14} & {\footnotesize{}20} & {\footnotesize{}22269} & {\footnotesize{}22276} & {\footnotesize{}406} & {\footnotesize{}1876}\tabularnewline
\hline 
{\footnotesize{}40.0} & {\footnotesize{}126} & {\footnotesize{}126} & {\footnotesize{}---} & {\footnotesize{}---} & {\footnotesize{}470} & {\footnotesize{}470} & {\footnotesize{}12} & {\footnotesize{}12} & {\footnotesize{}17325} & {\footnotesize{}17325} & {\footnotesize{}676} & {\footnotesize{}3486}\tabularnewline
\hline 
\end{tabular}
\par\end{centering}
\vspace{2bp}

\centering{}%
\begin{tabular}{|>{\centering}p{0.8cm}|>{\centering}p{0.8cm}|>{\centering}p{0.8cm}|>{\centering}p{0.8cm}|>{\centering}p{0.8cm}|>{\centering}p{0.8cm}|>{\centering}p{0.8cm}|>{\centering}p{0.8cm}|>{\centering}p{0.8cm}|>{\centering}p{0.8cm}|>{\centering}p{0.8cm}|>{\centering}p{0.8cm}|>{\centering}p{0.8cm}|}
\hline 
\multicolumn{13}{|c|}{(b) With Reinitiation GAL 1\% $k_{11}=0$}\tabularnewline
\hline 
aTc

{\footnotesize{}(ng ml$^{-1}$)} & {\small{}$G_{n}$to $G_{a}$}\\
{\small{}via $k_{1}$} & {\small{}$G_{a}$to $G_{n}$}\\
{\small{}via $k_{2}$} & {\small{}$G_{a}$to $G_{c}$}\\
{\small{}via $k_{3}$} & {\small{}$G_{c}$to $G_{a}$}\\
{\small{}via $k_{4}$} & {\small{}$G_{a}$to $G_{ar}$}\\
{\small{}via $k_{5}$} & {\small{}$G_{ar}$to $G_{a}$}\\
{\small{}via $k_{6}$} & {\small{}$G_{ar}$to $G_{r}$}\\
{\small{}via $k_{7}$} & {\small{}$G_{r}$to $G_{ar}$}\\
{\small{}via $k_{8}$} & {\small{}$G_{r}$to $G_{n}$}\\
{\small{}via $k_{9}$} & {\small{}$G_{n}$to $G_{r}$}\\
{\small{}via $k_{10}$} & {\small{}$G_{c}$to $M$}\\
{\small{}via $J_{m}$} & {\small{}$M$ to $P$}\\
{\small{}via $J_{p}$}\tabularnewline
\hline 
\hline 
{\footnotesize{}29.45} & {\footnotesize{}19} & {\footnotesize{}18} & {\footnotesize{}3716} & {\footnotesize{}3348} & {\footnotesize{}620} & {\footnotesize{}619} & {\footnotesize{}17} & {\footnotesize{}16} & {\footnotesize{}24497} & {\footnotesize{}24497} & {\footnotesize{}368} & {\footnotesize{}1980}\tabularnewline
\hline 
{\footnotesize{}31.56} & {\footnotesize{}20} & {\footnotesize{}24} & {\footnotesize{}5622} & {\footnotesize{}5052} & {\footnotesize{}551} & {\footnotesize{}555} & {\footnotesize{}9} & {\footnotesize{}13} & {\footnotesize{}22049} & {\footnotesize{}22054} & {\footnotesize{}570} & {\footnotesize{}3141}\tabularnewline
\hline 
{\footnotesize{}33.0} & {\footnotesize{}30} & {\footnotesize{}33} & {\footnotesize{}7581} & {\footnotesize{}6857} & {\footnotesize{}503} & {\footnotesize{}506} & {\footnotesize{}8} & {\footnotesize{}11} & {\footnotesize{}19636} & {\footnotesize{}19639} & {\footnotesize{}724} & {\footnotesize{}3753}\tabularnewline
\hline 
{\footnotesize{}34.73} & {\footnotesize{}38} & {\footnotesize{}41} & {\footnotesize{}11962} & {\footnotesize{}10845} & {\footnotesize{}562} & {\footnotesize{}566} & {\footnotesize{}6} & {\footnotesize{}10} & {\footnotesize{}14558} & {\footnotesize{}14562} & {\footnotesize{}1116} & {\footnotesize{}5673}\tabularnewline
\hline 
{\footnotesize{}36.0} & {\footnotesize{}43} & {\footnotesize{}48} & {\footnotesize{}10921} & {\footnotesize{}9892} & {\footnotesize{}420} & {\footnotesize{}425} & {\footnotesize{}4} & {\footnotesize{}9} & {\footnotesize{}15355} & {\footnotesize{}15361} & {\footnotesize{}1029} & {\footnotesize{}5252}\tabularnewline
\hline 
{\footnotesize{}40.0} & {\footnotesize{}68} & {\footnotesize{}68} & {\footnotesize{}15364} & {\footnotesize{}14026} & {\footnotesize{}244} & {\footnotesize{}245} & {\footnotesize{}3} & {\footnotesize{}4} & {\footnotesize{}9570} & {\footnotesize{}9571} & {\footnotesize{}1338} & {\footnotesize{}6648}\tabularnewline
\hline 
\end{tabular}\caption{\label{tab:Different-genetic-transitions table}Different genetic
transitions for different aTc concentrations}
\end{table}

We choose some random values of aTc and find out the different number
of transitions for a single cell with the help of a simulation based
on the Gillespie algorithm \cite{key-33}. It can be seen from table
\ref{tab:Different-genetic-transitions table}(b) that, the transitions
via $k_{3},k_{4},J_{m}$ and $J_{p}$ initially increase with aTc
and drop suddenly at aTc = 36 ng ml$^{-1}$. Whereas, transitions
via $k_{9}$ and $k_{10}$ initially decrease with aTc and undergoes
a sudden rise in number of transitions at aTc = 36 ng ml$^{-1}$.
It is to be noticed that although the variance of protein has a peak
around aTc = 36 ng ml$^{-1}$, due to the value of mean protein, the
maximum Fano factor (protein) is around aTc = 30 ng ml$^{-1}$ instead
of aTc = 36.0 ng ml$^{-1}$. At this point, on the basis of figure
\ref{fig:Variation-of-mean- and-variance}(a) and (b), one can ask
whether the mean values of mRNAs and proteins from reaction scheme
\ref{fig:Non-competetive network}(b) (i.e. with reinitiation ) can
always be greater than those obtained from the process \ref{fig:Non-competetive network}(a)
(i.e. without reinitiation) for any other parameters. We found the
answer to be in the negative. This has been explained in detail at
Appendix-D.

\section{Conclusions}

Regulation of gene expression and control of noise have many biological
and pharmacological significance \cite{key-37,key-38,key-56}. Transcription
factors (activators and repressors), that initiates the regulation
of gene expression, can act as a tumor suppressor in prostate cancer
\cite{key-37}. Stochastic gene expression fluctuations (i.e. noise)
are used to modulate reactivation of HIV from latency- a quiescent
state that is a major barrier to an HIV cure. Noise enhancers reactivates
latent cells, while noise suppressors stabilized latency \cite{key-56}. 

In this paper, we have studied a gene transcription regulatory architecture
observed in synthetic yeast GAL1$^{*}$ promoter. Activator and repressor
molecules bind the GAL1$^{*}$ promoter in a non-competitive fashion
to regulate the transcription. We focus on the reinitiation of transcription
by RNAP II as suggested by Blake \textit{et al.} \cite{key-03,key-06}.
In our recent works \cite{key-26,key-28}, we have found that the
Fano factor (at mRNA level) goes well below the sub-Poissonian region
and, after attaining a minima, it raises towards the super-Poissonian
regime crossing the Poissonian level. Now, we have made an attempt
to check whether the same nature of the Fano factor is valid in a
four-state process when both activator and repressor are in action.
We found the result in the positive. We have illustrated more effect
of trancriptional reinitiation on the non-competitive network along
with $k_{4}$ and aTc. We noticed that the Fano factor (protein) can
have a dip rather than a peak as observed in \cite{key-13}. 

Some aTc dependent features were observed by Blake \textit{et al.}
using reaction scheme \ref{fig:Non-competetive network}(b) in \cite{key-06}
and they have claimed that the noise in gene expression is promoter-specific.
We observed from figure (\ref{fig:Non-comp-3D-Mean-NS}) that both
the mean and the Fano factor vary with GAL and aTc; whereas, only
the Fano factor can be varied with aTc considering GAL as a parameter.
Figures \ref{fig:Non-Com-Mean-diff-gal-atc}(b) and \ref{fig:Non-Com-NS-aTc}(b)
show that the Fano factor can be varied with GAL though the mean protein
level (with reinitiation) is almost independent of GAL. Similar conclusion
was drawn in \cite{key-13} as well. But we have pointed out that
this conclusion is valid only when the reinitiation is involved (figure
\ref{fig:Non-Com-Mean-diff-gal-atc}b and \ref{fig:Non-Com-NS-aTc}b).

In \cite{key-06} the authors have shown the time dependence behavior
of noise with aTc concentration as a function of time in their network.
Whereas we have focussed on the dependence of noise on different reaction
rates. As we have included the extra path (fig. \ref{fig:Non-competetive network}c
and \ref{fig:Intermediate-state-consideration}b) of possible interactions
and reduced the approximations to possible extent, our network and
calculations became more complex and more general. In our analytical
computation, we have found expressions for mean and the Fano factor
at mRNA and protein levels. This can indeed be useful for further
rigorous study of non-competitive network via any of its parameters.
The analytical expression of the Fano factor with transcriptional
efficiency has been matched with the experimental data points of \cite{key-03}.
Our analytical curves and the simulation results are similar to those
of \cite{key-03}. Moreover, the experimental data points and the
analytical curve for the Fano factor (figure \ref{fig:Non-Com-mean}c)
at full aTc have a mismatch for lower transcriptional efficiency.
So we tried for the different set of rate constants to match the analytical
results and experimental data points of \cite{key-03}. Along with
the consideration of extra paths, the availability of analytical expressions
enables us to do that efficiently. We have found analytically a different
set of rate constants that matches the experimental data points of
mean and the Fano factor for all values of transcriptional efficiency.
We also observe that the choice of rate constants is not unique. There
can be other sets of rate constants that may give good fitting of
analytical curves with the experimental data points. The minimization
of relative error and mean squared error support the robustness of
our model estimations of parameters (see Appendix-E).

We checked how mean, Fano factor and variance behave in presence of
both activator and repressor. Among these two, repressor (bound by
aTc) has more governing role to the regulation of noise. There are
three regions of operation of aTc. Along with this, the reaction rates
$k_{4}$ and $k_{11}$ are also important nobs in noise regulation.
Also the factor \textit{e} appearing in the reaction rates, can play
a significant role in reducing noise in presence of highly active
aTc concentration.

We noticed that there are some anomalous nature of the Fano factor
of mRNA and variance of protein against aTc at lower GAL concentration
(<5\%). It is very interesting that although non-competitive network
has more noise than that of a two-state process, noise can be reduced
to sub-Poissonian region. But it can not be reduced below the level
presented by a two-state network with the help of transcriptional
reinitiation \cite{key-47}. We observed that although mean mRNA against
GAL or against aTc for a network with trancriptional reinitiation
is higher than that of a network without reinitiation but mean mRNA
can be both higher or lower when we plot them against other rate constants
(considering as variable).

For $k_{11}=0$ the reaction scheme \ref{fig:Non-competetive network}(c)
is same as \ref{fig:Non-competetive network}(b) that was proposed
by Blake \textit{et al.} \cite{key-03}. From analytical expressions
it has been shown that in the limit of aTc\textrightarrow \ensuremath{\infty}
and $k_{11}=0$ our model (reaction scheme \ref{fig:Non-competetive network}(c))
reduces to a well established two-state model with reinitiation producing
all the equations and plots as in \cite{key-26,key-28}. We have found
this can be achieved at aTc = 100 practically.

Along with this model, we conducted a parallel study of an activator-repressor
binding competitive network. We found a higher noise there in comparison
with the non-competitive binding. However the mean remains same for
the identical set of rate constants \cite{key-54}. A similar experimental
work was developed by Rossi \textit{et al.} \cite{key-55} and argued
that activator and repressor binding may work as a rheostat in putting
on/off a gene. Braichencho \textit{et al.} \cite{key-50} modeled
a three-state activator-repressor system where there is a proximal
promoter-pausing which can be effectively described by a two-state
model. Another competitive model has been proposed recently in \cite{key-53}
where the regulation of gene expression by TFs can occur via two cross
talking parallel pathways : basal and external signal.

Our proposed model may put an interest in synthetic biology, where
an engineering design approach is being studied to understand the
modified functionality of a biological system. From the analytical
approach, it is possible to predict the average and standard deviation
of the number of transcribed proteins. This model is useful to understand
the architecture of interactions which may buffer the stochasticity
inherent to gene transcription. There can be further extensive study
on the model considering the effect of other parameters that are not
included in present analysis.

\section*{\hspace{12em}Appendix - A}

In an attempt to deduce the expressions of mean and the Fano factors,
we use a moment generating function which is defined as,

\begin{equation}
F(z_{i},t)=\sum_{n=0}^{\infty}z_{i}^{n_{i}}p(n_{i},t)\label{eq:A1_generating fun}
\end{equation}
Here, $i=1,2,......,6$.

we have,

\begin{equation}
\begin{array}{ccc}
\frac{\partial F(z_{i},t)}{\partial t} & = & \sum_{n=0}^{\infty}z_{i}^{n_{i}}\frac{\partial p(n_{i},t)}{\partial t}\hspace{16bp}\hspace{16bp}\hspace{16bp}\hspace{24bp}\hspace{20bp}\hspace{16bp}\hspace{20bp}\hspace{20bp}\\
 & = & k_{1}(z_{1}-1)[lF-z_{1}\frac{\partial F}{\partial z_{1}}-z_{2}\frac{\partial F}{\partial z_{2}}-z_{3}\frac{\partial F}{\partial z_{3}}-z_{4}\frac{\partial F}{\partial z_{4}}]\\
 &  & +k_{2}(1-z_{1})\frac{\partial F}{\partial z_{1}}+k_{3}(z_{2}-z_{1})\frac{\partial F}{\partial z_{1}}+k_{4}(z_{1}-z_{2})\frac{\partial F}{\partial z_{2}}\\
 &  & +k_{5}(z_{3}-z_{1})\frac{\partial F}{\partial z_{1}}+k_{6}(z_{1}-z_{3})\frac{\partial F}{\partial z_{3}}+k_{7}(z_{4}-z_{3})\frac{\partial F}{\partial z_{3}}\\
 &  & +k_{8}(z_{3}-z_{4})\frac{\partial F}{\partial z_{4}}+k_{9}(1-z_{4})\frac{\partial F}{\partial z_{4}}\\
 &  & +k_{10}(z_{4}-1)[lF-z_{1}\frac{\partial F}{\partial z_{1}}-z_{2}\frac{\partial F}{\partial z_{2}}-z_{3}\frac{\partial F}{\partial z_{3}}-z_{4}\frac{\partial F}{\partial z_{4}}]\\
 &  & +k_{11}(1-z_{2})\frac{\partial F}{\partial z_{2}}+J_{m}(z_{1}z_{5}-z_{2})\frac{\partial F}{\partial z_{2}}+k_{m}(1-z_{5})\frac{\partial F}{\partial z_{5}}\\
 &  & +J_{p}(z_{6}-1)z_{5}\frac{\partial F}{\partial z_{5}}+k_{p}(1-z_{6})\frac{\partial F}{\partial z_{6}}
\end{array}\label{eq:A.2_generating fun moments}
\end{equation}

In steady state, $\frac{\partial F(z_{i},t)}{\partial t}=0$ and for
total probability, $F((z_{i}=1,0)=1$

Now, by setting {[}$\frac{\partial}{\partial z_{1}}(\frac{\partial F}{\partial t})]_{z_{i}=1}=0$,
we get $\frac{\partial F}{\partial z_{1}}=f_{1}(say)=\thinspace<n_{1}>\thinspace=$
average number of gene at state $G_{a}$.

similarly, by setting {[}$\frac{\partial}{\partial z_{1}}(\frac{\partial^{2}F}{\partial z_{1}\partial t})]_{z_{i}=1}=0$
will give $\frac{\partial^{2}F}{\partial z_{1}^{2}}=f_{11}(say)$
and so on. Proceeding in the same way we obtain ,

\[
f_{5}=\thinspace<n_{5}>\thinspace=mean\thinspace mRNA
\]

and 
\[
f_{6}=\thinspace<n_{6}>\thinspace=mean\thinspace Protein
\]
\[
Fano\thinspace factor\thinspace(mRNA)=\frac{variance\thinspace of\thinspace mRNA}{mean\thinspace mRNA}=\frac{f_{55}+f_{5}-f_{5}^{2}}{f_{5}}
\]
\[
Fano\thinspace factor\thinspace(Protein)=\frac{variance\thinspace of\thinspace Protein}{mean\thinspace Protein}=\frac{f_{66}+f_{6}-f_{6}^{2}}{f_{6}}
\]

\section*{\qquad{}\qquad{}\qquad{}\qquad{}\hspace{0em}\hspace{4em}Appendix
- B}

\qquad{} The parameters used in equation (\ref{eq:NonWR-M-2}),(\ref{eq:NC-FFm-3})
and (\ref{eq:NC-FFp-4}) are given below

\,

$A=-\frac{b_{19}J_{m}}{b_{20}k_{m}}$, $B=[\frac{J_{p}}{k_{m}+k_{p}}-\frac{J_{m}J_{p}k_{8}k_{3}^{2}(k_{10}(b_{8}-b_{5}k_{1})-b_{15}k_{1})}{b_{14}k_{p}(k_{m}+k_{p})}+\frac{b_{18}k_{8}(k_{m}+k_{1})}{(b_{8}-b_{4}k_{1})b_{14}(k_{m}+k_{p})}+\frac{b_{19}}{b_{20}}C]$,\\
 $C=\frac{b_{17}}{b_{14}(k_{m}+k_{p})}-\frac{J_{m}J_{p}}{k_{m}(k_{m}+k_{p})}-\frac{b_{1}b_{18}k_{8}}{(b_{8}-b_{4}k_{1})b_{14}(k_{m}+k_{p})}-\frac{(b_{2}+k_{m})}{k_{3}}(\frac{J_{m}J_{p}k_{3}^{2}(k_{5}(b_{8}-b_{5}k_{1})+b_{15}k_{8})}{b_{14}(k_{m}+k_{p})}+\frac{b_{18}(b_{13}-k_{8}k_{m})}{(b_{8}-b_{4}k_{1})b_{14}(k_{m}+k_{p})})$,

$b_{1}=J_{m}-k_{1}+k_{4}$,\hspace{6bp} $b_{2}=J_{m}+k_{4}+k_{11}$,\hspace{6bp}
$b_{3}=k_{8}+k_{9}+k_{10}$,\hspace{6bp} $b_{4}=k_{m}+k_{6}+k_{7}$, 

$b_{5}=k_{p}+k_{6}+k_{7}$,\hspace{6bp} $b_{6}=k_{1}+k_{2}+k_{3}+k_{5}$,\hspace{6bp}
$b_{7}=k_{8}(k_{10}-k_{7})$,\hspace{6bp} $b_{8}=(k_{6}-k_{1})k_{8}$,

$b_{9}=k_{1}(k_{7}k_{9}+k_{6}(k_{8}+k_{9})+k_{6}k_{8}k_{10})$,\hspace{6bp}
$b_{10}=k_{1}(k_{6}+k_{7}+k_{8})-k_{6}k_{8}$,

$b_{11}=k_{8}k_{10}+k_{7}(k_{9}+k_{10})+k_{6}b_{3}$,\hspace{6bp}
$b_{12}=k_{5}b_{3}-k_{8}k_{10}$,\hspace{6bp} $b_{13}=k_{1}k_{5}-b_{6}k_{8}$, 

$b_{14}=k_{3}((-b_{15}((b_{2}+k_{p})(b_{13}-k_{8}k_{p})+b_{1}k_{3}k_{8})+(k_{1}k_{p}+b_{10})(k_{3}k_{8}k_{10}-(b_{2}+k_{p})(b_{12}+k_{5}k_{p}))$,

$b_{15}=-b_{5}(b_{3}+k_{p})-b_{7}$,\hspace{10bp} $b_{16}=-b_{4}(b_{3}+k_{m})-b_{7}$,

$b_{17}=J_{m}J_{p}k_{3}(b_{15}(b_{13}-k_{8}k_{p})-(b_{8}-b_{5}k_{1})(b_{12}+k_{5}k_{p}))$, 

$b_{18}=J_{m}J_{p}k_{3}^{2}((b_{8}-b_{5}k_{1})(b_{3}+b_{4}+k_{p})-b_{15}k_{1})$,
\hspace{8bp}$b_{19}=k_{3}k_{8}(k_{10}(b_{8}-b_{4}k_{1})-(k_{m}+k_{1})b_{16})$, 

$b_{20}=(k_{3}(-b_{1}b_{16}k_{8}-k_{10}k_{8}(b_{8}-b_{4}k_{1}))+((b_{8}-b_{4}k_{1})(b_{12}+k_{5}k_{m})-b_{16}(b_{13}-k_{8}k_{m}))(b_{2}+k_{m}))$. 

\section*{\hspace{12em}\hspace{0em}Appendix - C}

Expression for mean mRNA and protein levels for transcription without
reinitiation are given by

\begin{equation}
m^{WTR}=\frac{a_{6}J_{m}}{(a_{6}+a_{5})k_{m}};\quad p^{WTR}=\frac{m^{WTR}\,J_{p}}{k_{p}}\label{eq:APP-1}
\end{equation}

where $a_{6}=\left(a_{1}k_{1}+k_{6}k_{8}k_{10}\right)$, $a_{5}=a_{1}k_{2}+a_{3}k_{2}+a_{2}k_{1}k_{5}+a_{4}$,
$a_{1}=k_{7}k_{9}+k_{6}(k_{8}+k_{9})$, $a_{2}=k_{7}+k_{8}+k_{9}$,
$a_{3}=k_{6}k_{10}+k_{7}k_{10}+k_{8}k_{10}$, $a_{4}=k_{5}k_{7}k_{10}+k_{5}k_{8}k_{10}+k_{5}k_{7}k_{9}$,

\,

The expression for the Fano factor at mRNA levels is given by

\begin{equation}
FF_{m}^{WTR}=1+\frac{g_{23}k_{8}J_{m}^{2}}{g_{20}k_{m}}+X-m^{WTR}\label{eq:APP-2}
\end{equation}

where $X=\frac{g_{22}k_{8}J_{m}^{2}\left(g_{20}\left(2g_{19}\left(g_{1}+k_{m}\right)\left(g_{3}\left(k_{m}+k_{1}\right)+k_{1}k_{10}\right)-g_{18}\left(g_{15}\left(k_{m}+k_{1}\right)+g_{10}k_{10}\right)\right)-g_{23}g_{24}\right)}{(g_{20}J_{m}(-g_{19}\left(g_{1}+k_{m}\right)\left(g_{13}-2g_{4}k_{8}J_{m}\right)+\left(g_{18}J_{m}\left(-g_{10}\left(g_{2}+k_{m}\right)-g_{15}k_{1}+g_{25}k_{8})\right)\right)+g_{20}g_{22}g_{24}k_{m})},$

$g_{25}=2g_{6}\left(g_{1}\left(k_{7}k_{8}-\left(k_{1}+k_{8}\right)k_{10}\right)-g_{5}k_{5}\right)-g_{8}g_{21}$,

$g_{24}=g_{18}J_{m}\left(g_{10}\left(k_{5}k_{m}-k_{8}k_{10}\right)+g_{15}\left(k_{1}k_{5}-k_{8}\left(g_{3}+k_{m}\right)\right)+g_{2}g_{10}k_{5}\right)+g_{19}\left(g_{13}k_{5}-g_{12}k_{8}\right)\left(g_{1}+k_{m}\right)$, 

$g_{23}=2g_{3}g_{16}\left(k_{m}+k_{1}\right)+g_{11}g_{18}\left(k_{m}+k_{1}\right)+k_{10}\left(2g_{16}k_{1}+g_{9}g_{18}\right)$,

$g_{22}=g_{18}\left(k_{8}\left(2g_{2}g_{6}\left(k_{7}-k_{10}\right)+2\left(g_{5}-g_{2}g_{4}\right)g_{21}\right)J_{m}^{2}-g_{14}J_{m}\right)-g_{16}\left(2g_{4}k_{8}J_{m}^{2}-g_{13}J_{m}\right)$,

$g_{21}=2\left(g_{1}^{2}+g_{3}g_{1}-g_{4}k_{5}\right)$, 

$g_{20}=g_{18}\left(k_{8}\left(g_{9}k_{10}J_{m}-g_{11}J_{m}\left(-g_{3}-k_{m}\right)\right)-g_{14}k_{5}\right)-g_{16}\left(g_{12}k_{8}-g_{13}k_{5}\right)$,

$g_{19}=J_{m}[k_{8}\left(g_{10}\left(k_{10}-k_{7}\right)-g_{4}g_{15}\right)-g_{10}\left(g_{2}+k_{m}\right)-g_{15}k_{1}]$,

$g_{18}=g_{17}k_{8}+g_{13}\left(g_{1}+k_{m}\right)$, \hspace{8bp}$g_{17}=2J_{m}(g_{3}g_{4}+g_{5})$, 

$g_{16}=k_{8}J_{m}\left(g_{9}\left(k_{10}-k_{7}\right)-g_{4}g_{11}\right)+g_{14}\left(g_{1}+k_{m}\right)$,\hspace{8bp}
$g_{15}=2g_{1}(g_{8}k_{5}-g_{6}k_{8}k_{10})$,

$g_{14}=J_{m}(g_{11}k_{1}+g_{9}\left(g_{2}+k_{m})\right)$, \hspace{8bp}$g_{13}=2k_{1}J_{m}(\left(-g_{2}-k_{m}\right)-g_{3})$,

$g_{12}=2J_{m}(g_{3}\left(-g_{3}-k_{m}\right)-k_{1}k_{10})$,\hspace{8bp}
$g_{11}=-g_{7}k_{5}-2g_{2}g_{6}k_{10}$,

$g_{10}=2g_{1}(g_{8}k_{8}-g_{6}\left(\left(g_{2}+g_{3}\right)k_{8}-k_{1}k_{5})\right)$,\hspace{8bp}
$g_{9}=2g_{6}\left(k_{1}k_{10}-g_{2}\left(g_{2}+g_{3}\right)\right)-g_{7}k_{8}$,

$g_{8}=2k_{8}g_{5}-2g_{1}\left(\left(g_{1}+g_{2}\right)k_{1}-g_{4}k_{8}\right)$,\hspace{8bp}
$g_{7}=4g_{1}\left(g_{5}-g_{2}g_{4}\right)$,\hspace{8bp} $g_{6}=2\left(g_{1}k_{1}-g_{4}k_{8}\right)$, 

$g_{5}=k_{1}(k_{7}-k_{10})$,\hspace{8bp} $g_{4}=(k_{6}-k_{1})$,\hspace{8bp}
$g_{3}=(k_{1}+k_{2}+k_{5})$, 

$g_{2}=(k_{8}+k_{9}+k_{10})$,\hspace{8bp} $g_{1}=(k_{6}+k_{7})$.

\,

The expression of the Fano factor at protein levels is given by

\begin{equation}
FF_{p}^{WTR}=1+\frac{J_{p}}{k_{m}+k_{p}}-\frac{k_{1}k_{8}J_{m}J_{p}}{h_{1}k_{p}\left(k_{m}+k_{p}\right)}+\frac{h_{4}\left(h_{1}k_{8}k_{10}-h_{2}k_{1}k_{8}\right)J_{m}J_{p}}{h_{1}h_{8}k_{p}\left(k_{m}+k_{p}\right)}+Y+Z-p^{WTR}\label{eq:APP-3}
\end{equation}

where

$Y=\frac{k_{8}J_{m}^{3}J_{p}\left(h_{11}-g_{24}\left(k_{10}\left(2g_{16}k_{1}+g_{9}g_{18}\right)+h_{7}\left(k_{m}+k_{1}\right)\right)\right)(h_{3}k_{1}\left(g_{13}+2\left(k_{1}-k_{6}\right)k_{8}J_{m}\right)+\frac{h_{9}h_{10}}{g_{20}k_{m}}-h_{4}\left(2h_{6}\left(k_{1}-k_{6}\right)k_{8}J_{m}+h_{5}\right))}{g_{18}h_{8}\left(k_{m}+k_{p}\right)\left(g_{20}J_{m}\left(g_{18}J_{m}\left(-g_{10}\left(g_{2}+k_{m}\right)-g_{15}k_{1}+g_{25}k_{8}\right)-g_{19}\left(g_{1}+k_{m}\right)\left(2g_{4}k_{8}J_{m}+g_{13}\right)\right)+g_{22}g_{24}\right)},$

$Z=\frac{2k_{8}\left(h_{3}k_{1}-h_{4}h_{6}\right)J_{m}^{2}J_{p}\left(g_{3}k_{m}+k_{1}\left(g_{3}+k_{10}\right)\right)}{g_{18}h_{8}\left(k_{m}+k_{p}\right)}+\frac{h_{9}k_{8}J_{m}^{2}J_{p}\left(k_{1}\left(2g_{16}k_{10}+h_{7}\right)+g_{9}g_{18}k_{10}+h_{7}k_{m}\right)}{g_{18}g_{20}h_{8}k_{m}\left(k_{m}+k_{p}\right)},$

$h_{11}=g_{20}\left(2g_{19}\left(g_{1}+k_{m}\right)\left(g_{3}k_{m}+k_{1}\left(g_{3}+k_{10}\right)\right)-g_{18}\left(g_{15}k_{m}+g_{15}k_{1}+g_{10}k_{10}\right)\right)$,

$h_{10}=2k_{8}\left(g_{16}\left(k_{1}-k_{6}\right)+g_{18}\left(g_{2}g_{6}k_{7}-g_{2}g_{6}k_{10}-g_{2}g_{4}g_{21}+g_{5}g_{21}\right)\right)J_{m}+g_{13}g_{16}-g_{14}g_{18}$, 

$h_{9}=k_{m}\left(h_{4}\left(h_{5}k_{5}-g_{12}h_{6}k_{8}\right)-h_{3}\left(g_{18}k_{8}+k_{1}\left(g_{13}k_{5}-g_{12}k_{8}\right)\right)\right)+g_{18}h_{8}$,

$h_{8}=(h_{1}h_{3}-h_{2}h_{4})$, \hspace{10bp}$h_{7}=2g_{3}g_{16}+g_{11}g_{18}$,

$h_{6}=g_{1}+g_{2}+k_{m}+k_{p}$,

$h_{5}=(k_{m}+k_{p}+g_{1}+g_{2})g_{13}-g_{18}$,

$h_{4}=g_{4}k_{8}-k_{1}\left(g_{1}+k_{p}\right)$,\hspace{10bp} $h_{3}=k_{8}(k_{7}-k_{10})-(g_{1}+k_{p})(g_{2}+k_{p})$,

$h_{2}=k_{5}(g_{2}+k_{p})-k_{8}k_{10}$,\hspace{10bp} $h_{1}=k_{1}k_{5}-k_{8}(g_{3}+k_{p})$

\section*{\hspace{12em}Appendix - D}

In figure \ref{fig:Variation-of-mean- and-variance}(a) and (b) it
is seen that mean mRNA and mean protein in case of transcriptional
reinitiation based network is greater than that of without reinitiation
network. But the mean values against other variables like $J_{m}$
shows that $m^{WR}$ can be high or less than $m^{WTR}$ as shown
in figure below.

\begin{figure}[H]
\begin{centering}
\includegraphics[width=5.5cm,height=3.5cm]{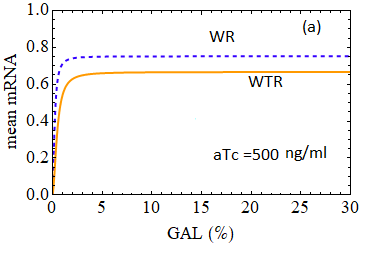} \includegraphics[width=5.5cm,height=3.5cm]{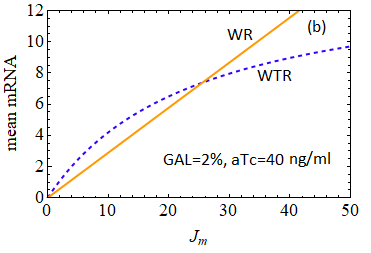}
\par\end{centering}
\caption{\label{fig:mean WR WTR}Variation of mean mRNA against GAL keeping
aTc = 500 ng ml$^{-1}$ as parameter in (a) and mean mRNA vs $J_{m}$
with aTc = 40 ng ml$^{-1}$ and GAL = 2\% as parameter in (b) while
other rate constants are chosen from Blake \textit{et al.} \cite{key-03}.}
\end{figure}

Figure \ref{fig:Variation-of-mean- and-variance}(a) and figure \ref{fig:mean WR WTR}(a)
shows that $m^{WR}$ is higher than $m^{WTR}$ keeping GAL and aTc
as parameter respectively but figure \ref{fig:mean WR WTR}(b) shows
a different scenario where we keep both GAL and aTc fixed. It is verified
that, the slope of mean mRNA curve becomes high against $J_{m}$ for
a higher concentration of GAL keeping aTc fixed. On the other hand,
if we increase aTc for a fixed GAL the slope of the curve goes high
against $J_{m}.$ Imposing the condition $m^{WR}=m^{WTR}$ we have
found a critical value of $J_{m}$ given by

\begin{equation}
J_{m}^{cr}=\frac{k_{3}\left(B_{7}-b_{11}k_{11}\right)-\left(b_{9}+B_{7}\right)k_{r}}{b_{9}+B_{7}}\label{eq:critical Jm for meanM}
\end{equation}

where, $B_{7}=a_{2}k_{1}k_{5}+a_{4}+b_{11}k_{2}$ and $k_{r}=k_{4}+k_{11}$

other parameters are supplied earlier.

\section*{\hspace{12em}Appendix - E}

\begin{figure}[H]

\centering{}\includegraphics[width=10cm,height=6cm]{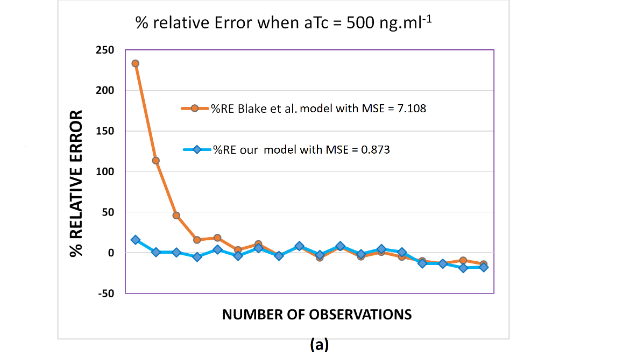}\includegraphics[width=8cm,height=6cm]{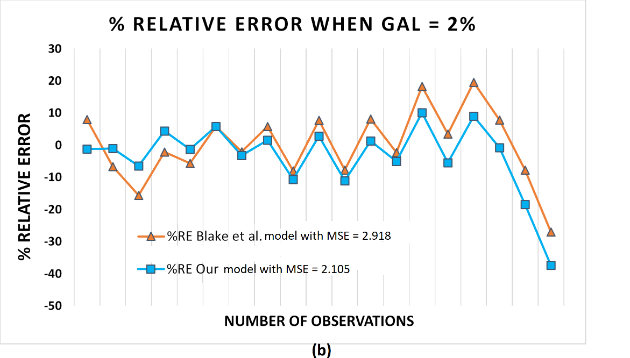}\caption{Relative Error for\label{fig:relative-error} : (a) fitting of curves
when aTc is fixed at 500 ng/ml. Blake \textit{et al. }\cite{key-03}
model offers MSE = 7.108 while our proposed model fits quite nicely
with the experimentally observed data with a minimize MSE = 0.873\protect \\
 (b) fitting of curves when aTc is fixed at 500 ng/ml. Blake \textit{et
al. }\cite{key-03} model offers MSE = 2.918 while our proposed model
fits better with the experimentally observed data with a reduced MSE
= 2.105}
\end{figure}

\textbf{Model fitting parameters: }We proposed an analytical model
that fits very much to the experimental data as supplied by Blake
\textit{et al.} \cite{key-03}. The parameters are chosen from the
supplementary material of \cite{key-03}. The form of parameters $k_{1}$
and $k_{2}$ as functions of GAL are determined from an exact analytical
treatment as described in the main text. While for the best estimates
of other parameters are revised through trial and error to minimize
the sum of squared error (SSE) and the mean square error (MSE). We
also show the relative percentile error (RE), in figure \ref{fig:relative-error}
(a) and (b) for the fitting of each of the data points, which is given
by 
\begin{equation}
RE=\frac{y_{a}-y_{e}}{y_{e}}.100\%
\end{equation}

where, \textit{a} stands for analytical and \textit{e} stands for
experimental.

\textbf{Uncertainty in fitted parameters :} The uncertainty of the
parameter estimates, is generally expressed by the mean square errors,
is proportional to the SSE and inversely proportional to the square
of the sensitivity coefficient of the model parameters \cite{key-58}.
The mean square fitting error is
\begin{equation}
\sigma^{2}=\frac{1}{n-k}\sum_{i=1}^{n}(y_{a}-y_{e})^{2}=\frac{Sum\thinspace of\thinspace squared\thinspace error}{(n-k)}
\end{equation}
\textit{n} is the number of observations and \textit{k} is the number
if parameters being determined. 

The sensitivity (\ensuremath{\mathscr{S}}) of a function $f(k)$ over
the parameter $k$ is given by
\begin{equation}
\mathscr{S}=\frac{k}{f(k)}.\frac{\partial f(k)}{\partial k}
\end{equation}

We obtain the sensitivity of the Fano factor (protein) over the fitting
parameters and calculate MSE of each parameter keeping others as constant.
\begin{equation}
MSE=\frac{\sigma^{2}}{\sum_{i=1}^{n}[\frac{\partial f(k)}{\partial k}]^{2}}
\end{equation}

where the denominator is the coefficient of sensitivity, squared and
summed over all observations.
\begin{center}
\begin{figure}[H]
\centering{}\includegraphics[width=12cm,height=8cm]{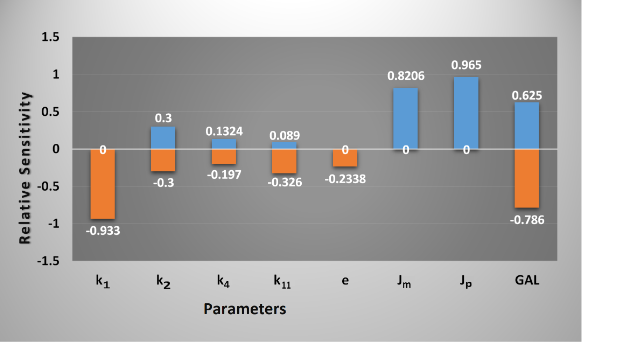}\caption{Relative Sensitivity of the Fano factor as functions of fitting parameters\label{fig:relative-sensitivity}}
\end{figure}
\par\end{center}

\begin{figure}[H]
\begin{centering}
\includegraphics[width=12cm,height=6cm]{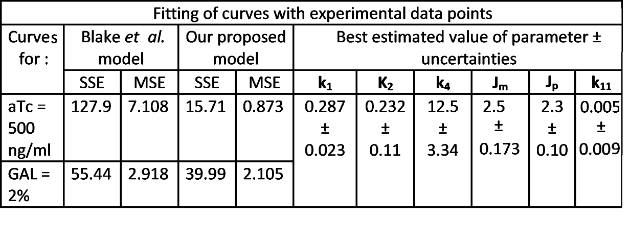}
\par\end{centering}
\begin{centering}
\caption{\label{fig:Table-for-the estimation_uncertainty}Table for the errors
in curve fitting, estimated parameter values and their uncertainties. }
\par\end{centering}
\end{figure}

The GAL is the most sensitive parameter and \textit{e} is the least
in order. We found that, the Fano factor is sensitive within a small
range of values of these parameters and with the best parameter estimation
and minimization of errors (see figure \ref{fig:Table-for-the estimation_uncertainty})
support the robustness of our result. The square root of the MSE is
the standard deviation, and the approximate 95\% confidence interval
for \textit{k} is \cite{key-58}

\begin{equation}
[k]_{95\%}=\kappa\pm2\surd MSE
\end{equation}

$\kappa$ is the best estimate value of parameter $k.$

\section*{Acknowledgment }

The author would like to acknowledge the helpful suggestions of Dr.
Rajesh Karmakar during the initial stage of the project, before he
deceased on 09th June 2021. The author also thanks Dr. Indrani Bose
and Dr. Arindam Lala for their valuable suggestions and discussions
on the paper.

\section*{Glossary}
\begin{itemize}
\item \textbf{Transcription factor : }Transcription factors (TFs) are proteins
which have DNA binding domains with the ability to bind to the specific
sequences of DNA (called promoter). They controls the rate of transcription.
If they enhance transcription they are called activators and termed
as repressors if inhibit transcription.
\item \textbf{Fano factor and Noise strength : }The Fano factor is the measure
of deviations of noise from the Poissonian behavior and is defined
as \cite{key-50,key-51} ,
\end{itemize}
\[
Fano\thinspace factor=\frac{variance}{mean}=\frac{(standard\thinspace deviation)^{2}}{mean}
\]

So, for a given mean, smaller the Fano factor implies smaller variance
and thus less noise. Therefore, the Fano factor gives a measure of
noise strength which is defined (mathematically) as \cite{key-02},

\[
noise\thinspace strength=\frac{variance}{mean}=\frac{(standard\thinspace deviation)^{2}}{mean}
\]

\begin{itemize}
\item \textbf{Transcriptional efficiency : }Transcriptional efficiency is
the ratio of instantaneous transcription to the maximum transcription.
\end{itemize}

\end{document}